\documentclass{pas}

\usepackage{multirow}
\usepackage[normalem]{ulem} 
\definecolor{darkgreen}{rgb}{0,0.5,0} 

\usepackage{bm}

\begin{document}

\lefttitle{Publications of the Astronomical Society of Australia}
\righttitle{Arun et al.}

\jnlPage{1}{??}        
\jnlDoiYr{2026}        
\doival{10.1017/pasa.xxxx.xx}

\articletitt{Research Paper}

\title{ROCKETS I: Investigating the Impact of the Rocket Effect on Two Nearby Young Open Clusters Encased in Infrared Bubbles}

\author{
\sn{Arun} \gn{Roy}$^{1,2,3,4}$,
\sn{Gopinathan} \gn{Maheswar}$^{4}$,
\sn{Subramaniam} \gn{Annapurni}$^{4}$,
\sn{Saha} \gn{Piyali}$^{5,6}$,
\sn{Sudheesh} \gn{T.~P.}$^{1,7,8}$,
\sn{Shridharan} \gn{B.}$^{9}$,
\sn{Mathew} \gn{Blesson}$^{1}$,
and
\sn{Akhil} \gn{K.~R.}$^{4}$
}

\affil{
$^{1}$Centre of Excellence in Astronomy and Astrophysics, Department of Physics and Electronics, CHRIST (Deemed to be University), Bangalore 560029, India\\
$^{2}$Instituto de Estudios Astrofísicos, Facultad de Ingeniería y Ciencias, Universidad Diego Portales, Av. Ejército Libertador 441, Santiago, Chile\\
$^{3}$Millennium Nucleus on Young Exoplanets and their Moons (YEMS), Chile\\
$^{4}$Indian Institute of Astrophysics, Sarjapur Road, Koramangala, Bangalore 560034, India\\
$^{5}$Academia Sinica Institute of Astronomy and Astrophysics, No.~1, Sec.~4, Roosevelt Road, Taipei 10617, Taiwan\\
$^{6}$National Astronomical Observatory of Japan, National Institutes of Natural Sciences, 2--21--1 Osawa, Mitaka, Tokyo 181--8588, Japan\\
$^{7}$St. Joseph's College, Moolamattom, Idukki, 685591, India\\
$^{8}$Department of Physics, Newman College, Thodupuzha 685585, India\\
$^{9}$Tata Institute of Fundamental Research, Homi Bhabha Road, Mumbai 400005, India
}

\corresp{R. Arun, Email: arunroyon@gmail.com}

\citeauth{
Arun R., Gopiathan M., Subramaniam A., Saha P., Sudheesh T.~P., Shridharan B., Mathew B., Akhil K.~R.,
ROCKETS I: Investigating the Impact of the Rocket Effect on Two Nearby Young Open Clusters Encased in Infrared Bubbles,
{\it Publications of the Astronomical Society of Australia},
{\bf 00}, 1--?? (2026).
https://doi.org/10.1017/pasa.xxxx.xx
}

\history{(Received xx xx xxxx; revised xx xx xxxx; accepted xx xx xxxx)}

\begin{abstract}
We introduce \textsc{ROCKETS} (Rocket-driven Cluster Kinematics \& Triggered Star-formation), a \textit{{\it Gaia}} based programme investigating kinematic fingerprints of stellar feedback or rocket effect in young star forming regions. We analyse two young clusters with O-type stars enclosed by mid-infrared bubbles, Collinder~69 and IC~1396 as a methodological demonstration on two nearby benchmark regions. Co-moving candidates are selected within an astrometric ellipse defined by the median and median absolute deviation of Class~I/II YSOs with {\it Gaia} data. For each star, we compute a relative proper-motion angle (RPMA) relative to the ionising source. In both regions the RPMA distribution shows a strong excess at RPMA~$<15^{\circ}$, corresponding to sources moving away from the respective ionising sources. We quantify this using the Rocket Effect Index (REI), finding  $\mathrm{REI}\simeq 1.1$ for Collinder~69 and $\mathrm{REI}\simeq 1.7$ for IC~1396. When considering only Class~I/II YSOs, the REI of both regions increases to $\mathrm{REI}\gtrsim 10$. Velocity-structure analysis of the sample shows that the RPMA~$<15^\circ$ subset has stronger positive pairwise expansion than the full co-moving population, providing additional support for the rocket effect. The outward-moving stars are not uniformly distributed; instead, their position angles are strongly anisotropic and preferentially aligned towards the bright-rimmed clouds. {\it Gaia} colour--magnitude diagrams show no strong age offset between sources moving outward and literature cluster members. The observed kinematic signatures support feedback-driven acceleration of gas and triggered star formation preferentially towards higher density regions of expanding bubbles due to the rocket effect.
\end{abstract}

\begin{keywords}
proper motions -- stars: formation –- stars: kinematics and dynamics –- open clusters and associations: general -- stars: massive
\end{keywords}

\maketitle

\section{Introduction}

Stellar feedback from massive stars plays a central role in regulating star formation and shaping the evolution of galaxies \citep{McKee2007ARA&A..45..565M,Kennicutt2012ARA&A..50..531K}. As massive ($\gtrsim$ 8 M$\odot$) stars are born within giant molecular clouds (GMCs), they simultaneously influence their natal environment through various feedback processes such as stellar winds, ionising radiation, radiation pressure, and supernova explosions \citep{Weaver1977ApJ...218..377W,Girichidis2020SSRv..216...68G}. These processes can either trigger further star formation (positive feedback) or inhibit it (negative feedback), depending on the local and global conditions \citep{Dale2015,Hopkins2014MNRAS.445..581H}.

Young star clusters represent dynamic environments where the interplay of stellar feedback mechanisms shapes both the morphology and kinematics of their parent molecular clouds. Among these processes, the influence of massive O-type stars is particularly significant. Through intense ultraviolet (UV) radiation and stellar winds, these stars ionise surrounding gas, creating HII regions and driving the formation of infrared (IR) bubbles, detectable at mid-IR wavelengths \citep{Churchwell2006ApJ...649..759C}. These feedback processes can trigger secondary star formation through mechanisms like Radiation-Driven Implosion (RDI; \citealp{Bertoldi1989ApJ...346..735B,Sugitani1989ApJ...342L..87S,Sugitani1991ApJS...77...59S}), and formation by collect and collapse in the peripheries of the expanding IR bubbles \citep{Elmegreen1977ApJ...214..725E,Dale2007MNRAS.375.1291D}.

The rocket effect (RE), conceptualised by \cite{Oort1955ApJ...121....6O} and \cite{Kahn1954BAN....12..187K}, occurs when the ionisation front associated with massive stars expands and interacts with the surface of a dense molecular clump. The intense UV radiation ionises the gas, driving a photoevaporative flow off the illuminated face. By conservation of momentum, the clump recoils and accelerates away from the ionising source, often producing bright rims and cometary morphologies pointing towards the ionising source \citep{Lefloch1994A&A...289..559L}. The resulting compression at the ionisation front can also promote collapse, so newly formed stars may inherit the motion same as the clumps \citep{Dale2015}. This effect is particularly relevant in regions near bright-rimmed clouds (BRCs), where dense gas clumps at the edges of HII regions are compressed, potentially initiating gravitational collapse and triggering star formation \citep{Sugitani1994ApJS...92..163S}. Observational studies have documented this effect in individual BRCs within larger IR bubbles (e.g. \citealp{Arun2021MNRAS.507..267A,Sahaa2022MNRAS.510.2644S,Saha2022MNRAS.515L..67S}). However, a systematic exploration of the RE’s role in shaping stellar kinematics and cluster morphology across multiple photon directions from ionising stars remains underexplored. Determining the RE can offer a means to estimate how efficiently stellar feedback triggers star formation and to trace the extent of positive feedback within evolving HII regions \citep{Dale2007MNRAS.375.1291D, 2015MNRAS.452.2794W}.

{\it Gaia}, a cornerstone mission of the European Space Agency (ESA), aims to create the most precise 3D map of our galaxy by surveying over a billion stars \citep{Gaia2016A&A...595A...1G}. The precise astrometric data from the subsequent data releases of {\it Gaia} mission \citep{Gaia2016A&A...595A...1G,Gaia2023A&A...674A...1G} has revolutionised our ability to study the dynamics of young stellar populations. By leveraging {\it Gaia} Early Data Release 3 (EDR3), and subsequent Data Release 3 (DR3), it is now possible to investigate the proper motions and spatial distributions of stars with unprecedented accuracy. These data provide a unique opportunity to investigate the kinematics of stars in and around young open clusters containing central O-type stars, where the RE can propel clouds radially away from the ionising source. If star formation is triggered by this effect, newly formed young stellar objects (YSOs) should generally share the motion of the cloud as it accelerates away from the ionising source  \citep{Dale2015,Sahaa2022MNRAS.510.2644S}. In this study, ROcket-driven Cluster Kinematics \& Triggered Star-formation (ROCKETS), we examine the kinematic signatures of the RE in two nearby young open clusters, Collinder 69 and IC 1396. These clusters were selected for their prominent IR bubble morphologies and central O-type stars, which are the primary drivers of stellar feedback. We used {\it Gaia} EDR3 astrometric data to identify stellar populations that are co-moving (share similar distance and proper motions) with the known members of the two open clusters. These populations extend up to the boundaries of the associated IR bubbles. We then analysed their proper motion distributions relative to the ionising sources. By calculating relative proper motion angles (RPMAs; \citealp{Sahaa2022MNRAS.510.2644S}) and with various statistical analysis, we quantify the extent to which the RE shapes the dynamics of stars in these clusters.

This paper is structured as follows: in Section 2, we describe the data selection and astrometric analysis. Section 3 presents the results and in Section 4, we discuss the implications of our findings for star formation processes, followed by a summary in Section 5.

\section{Data \& Analysis}

In this section, we describe the identification of the young clusters for ROCKETS, the selection of the two clusters analysed in this study, and the astrometric analysis using \textit{{\it Gaia}} data.

\subsection{Young Cluster Identification}

To study the effect of ionising stars on IR bubbles, we selected a sample of young open clusters (OCs) exhibiting clear IR-bubble morphology and containing central hot ionising O-type stars. The restriction to clusters with O-type stars is motivated by two considerations: the physics of the rocket effect itself and the astrometric requirements of our method. Since the RPMA is defined relative to the ionising source, the presence of a dominant central O-type star is essential. We started from the catalogue of 1222 open clusters compiled by \citet{Cantat2018A&A...618A..93C}. Using the cluster radius containing 50\% of the listed members ($r_{50}$) and the central coordinates, we cross-matched with the SIMBAD database and identified 95 clusters hosting at least one O-type star within $r_{50}$.

The presence of an IR bubble was then assessed by visual inspection of the 12\,$\mu$m W3 band images from the Wide-field Infrared Survey Explorer (WISE; \citealt{Wright2010AJ....140.1868W}), which traces polycyclic aromatic hydrocarbon (PAH) emission at the bubble rims. This step yielded 24 clusters with well-defined IR bubble structures; these 24 systems form the parent sample of the full ROCKETS programme.

This first paper (ROCKETS\,I) is a pilot study that focuses on the two nearest clusters (Collinder\,69 and IC\,1396, both at distances $\lesssim 1$\,kpc) with highly symmetric bubble morphologies. These two systems were selected as benchmark systems because their proximity ensures the highest \textit{Gaia} astrometric precision, while their geometric symmetry provides a clean reference frame for measuring radial motions relative to the central ionising source. A flow-chart summarising the complete selection process is shown in Figure~\ref{fig:flow}. Details of both clusters are given below.
\begin{figure}
    \includegraphics[width=\columnwidth]{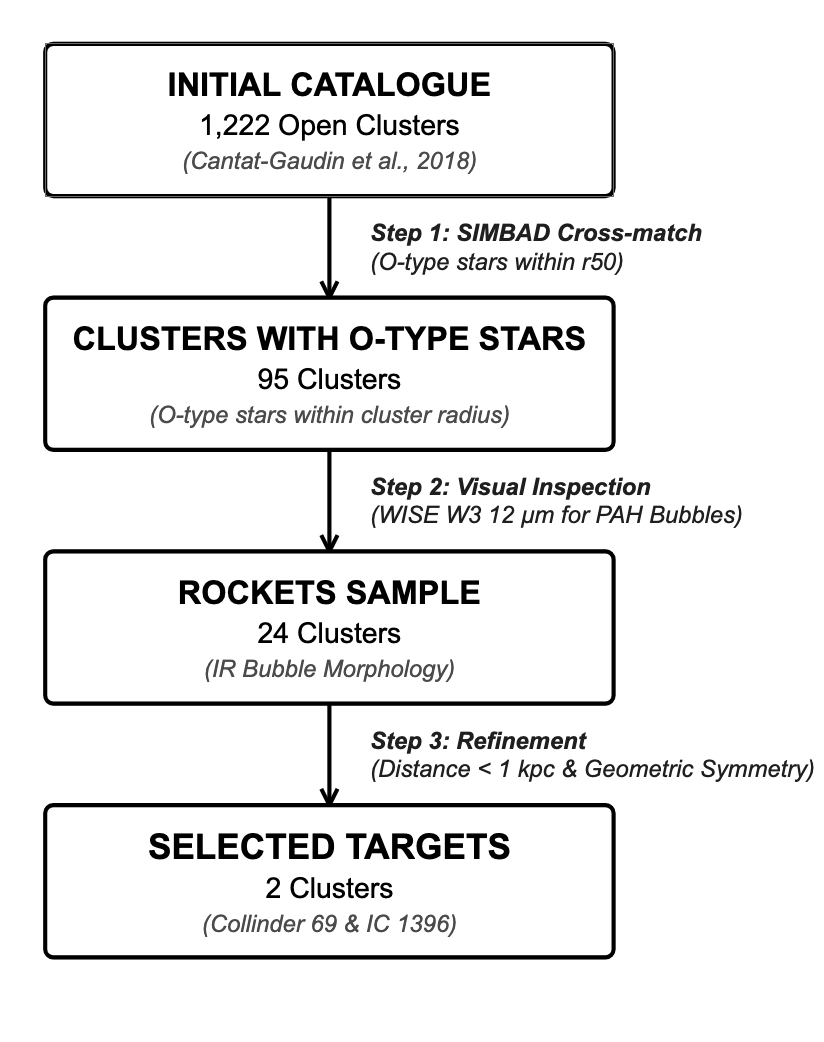}
 \caption{A schematic illustration of the selection criteria applied to the initial open cluster catalogue to identify the final two targets in this study.
    }
    \label{fig:flow}
\end{figure}

\subsubsection{Collinder 69}

Collinder 69 is a part of the $\lambda$ Orionis star forming region. The cluster has a relatively low extinction value $A_{V} \sim$ 0.36 mag \citep{1982duerr, 2008bayo}. The brightest star in the cluster is the O-type binary star, $\lambda$ Orionis with spectral type  O8IIIf+B0.5V \citep{1971conti,2011sota}. The cluster also hosts at least a dozen B type stars \citep{1977murdin, 1982duerr}. The age of the cluster is estimated to be $\sim$ 5 Myr \citep{2004barrado, 2011bayo, 2023healy} and the distance to the cluster is $\sim$ 400 pc \citep{1977murdin, 2011bayo, 2023healy}. Recently, two studies have demonstrated the RE towards BRC 17 and 18 \citep{Saha2022MNRAS.515L..67S,Sahaa2022MNRAS.510.2644S}, which are at the boundaries of the bubble morphology seen near Collinder 69. Unfortunately, $\lambda$ Orionis has a high RUWE value of 4.8, which makes it unreliable for the calculation of RPMA. Thus, we will use the weighted median (WM) of proper motion in Right Ascension ($\mu_{\alpha *}$) and proper motion in Declination ($\mu_{\delta}$) of the known cluster members for the calculation. Considering the mass segregation in clusters, using weights according to the brightness as prescribed by \cite{Arun2021MNRAS.507..267A} makes it a better proxy to the central O-type star.

The radius of the cluster given by \cite{Cantat2018A&A...618A..93C} is 0.99\textdegree. We intend to study stars that are formed in the expanding IR bubble. To determine these outer boundaries, WISE W3 band image, which traces PAHs was used, and a new radius was estimated. The estimate shows that the IR bubble extends up to 5.8\textdegree. The BRCs SFO 17 and 18 have a separation of 2.8\textdegree~from the cluster centre. Therefore, in all subsequent analyses, we consider only sources located within a 5.8\textdegree\ radius from the cluster centre.

\subsubsection{IC 1396}
The IC 1396 region is part of the large, star-forming Cepheus bubble \citep{Patel1998ApJ...507..241P}, ionised by the multiple system (O5+O9) HD 206267 \citep{Peter2012A&A...538A..74P, 2020maiz}. The central cluster of IC 1396, also known as Trumpler 37 \citep{Patel1995ApJ...447..721P, Patel1998ApJ...507..241P}, has a distance of 945 pc based on {\it Gaia} DR2 \citep{Sicilia2019A&A...622A.118S}. The region is estimated to have an age of $\sim$ 4 Myrs and an A\textsubscript{V} = 1.40 mag \citep{Pelayo2023A&A...669A..22P}. The region contains the well-known BRCs SFO 36, 37, 38, and 39, with the central ionising source HD 206267 believed to be responsible for their ionisation \citep{Sugitani1991ApJS...77...59S}. Similar to Collinder 69, the central star of IC 1396, HD 206267 also has a high RUWE value of 5.074. Thus, we use WM values of the cluster members for the calculation of RPMA. The radius of the IC 1396 cluster given by \cite{Cantat2018A&A...618A..93C} is 0.3\textdegree. But the WISE 12 $\mu$m W3 image shows that the IR bubble extends up to 1.4\textdegree. Our analysis includes sources within the bubble and near the surrounding BRCs, providing a more complete picture of the RE.

\subsection{{\it Gaia} Astrometric Data}

{\it Gaia} DR3 includes comprehensive data on stellar positions, distances, proper motions, and various stellar parameters, significantly enhancing our understanding of stellar and galactic evolution. For our analysis of young open clusters, we use the {\it Gaia} EDR3 astrometric data. The current study does not use radial velocities, thus all the astrometric data taken are from  {\it Gaia} Early Data Release 3 (EDR3).

We note that direct conversion of {\it Gaia} parallax values may lead to inherent issues \citep{bailerjohns2018AJ....156...58B,bailerjohns2021AJ....161..147B}. Therefore, in this work we do not estimate distances directly from the parallaxes. Instead, we use the distances tabulated by \citet{bailerjohns2021AJ....161..147B}, who provide both geometric and photogeometric distance estimates for {\it Gaia} EDR3 sources. Since the clusters analysed here are located within $\sim$1\,kpc, we adopted the geometric distances from that catalogue for all sources in our sample.

\begin{figure}
    \includegraphics[width=\columnwidth]{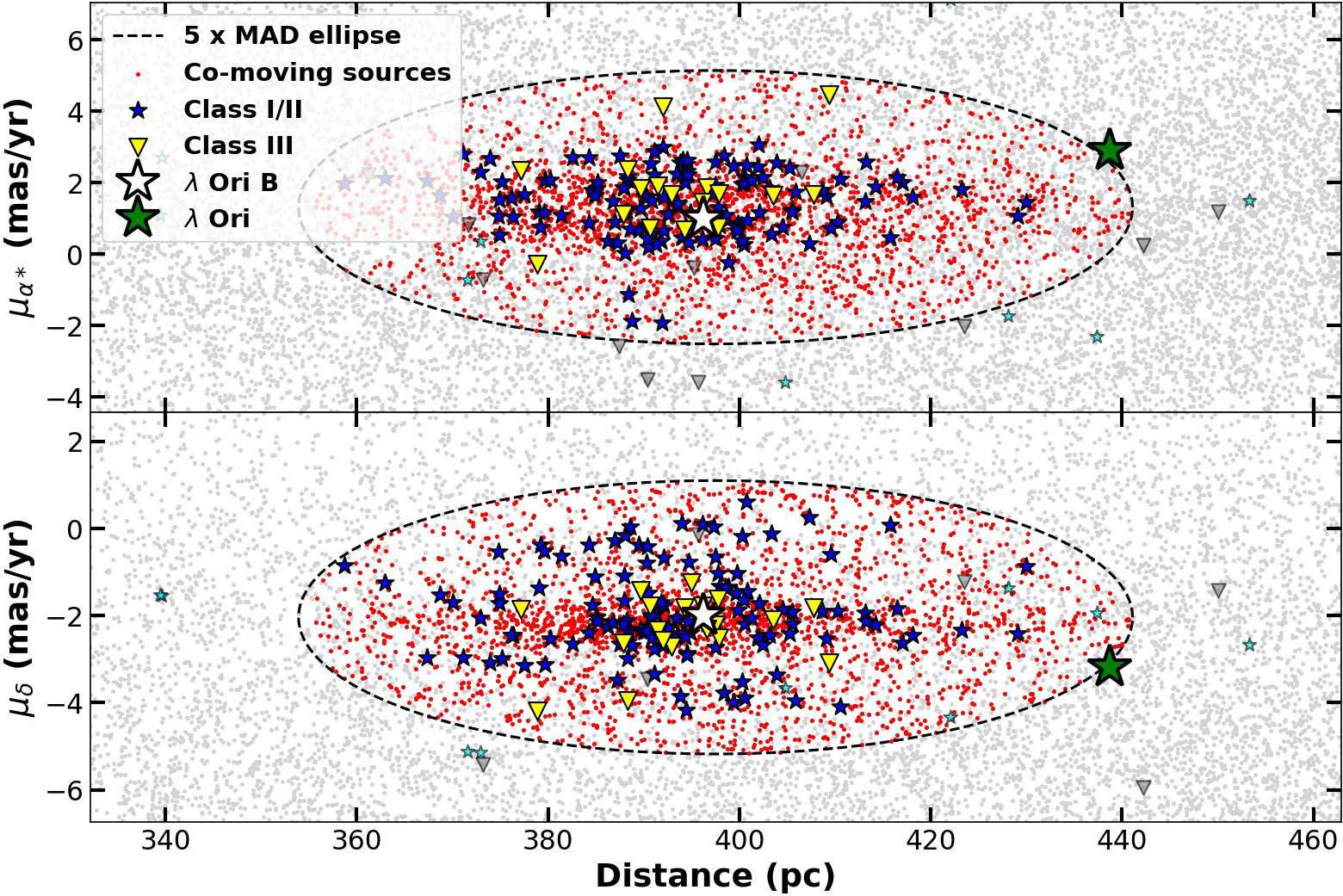}
    \includegraphics[width=\columnwidth]{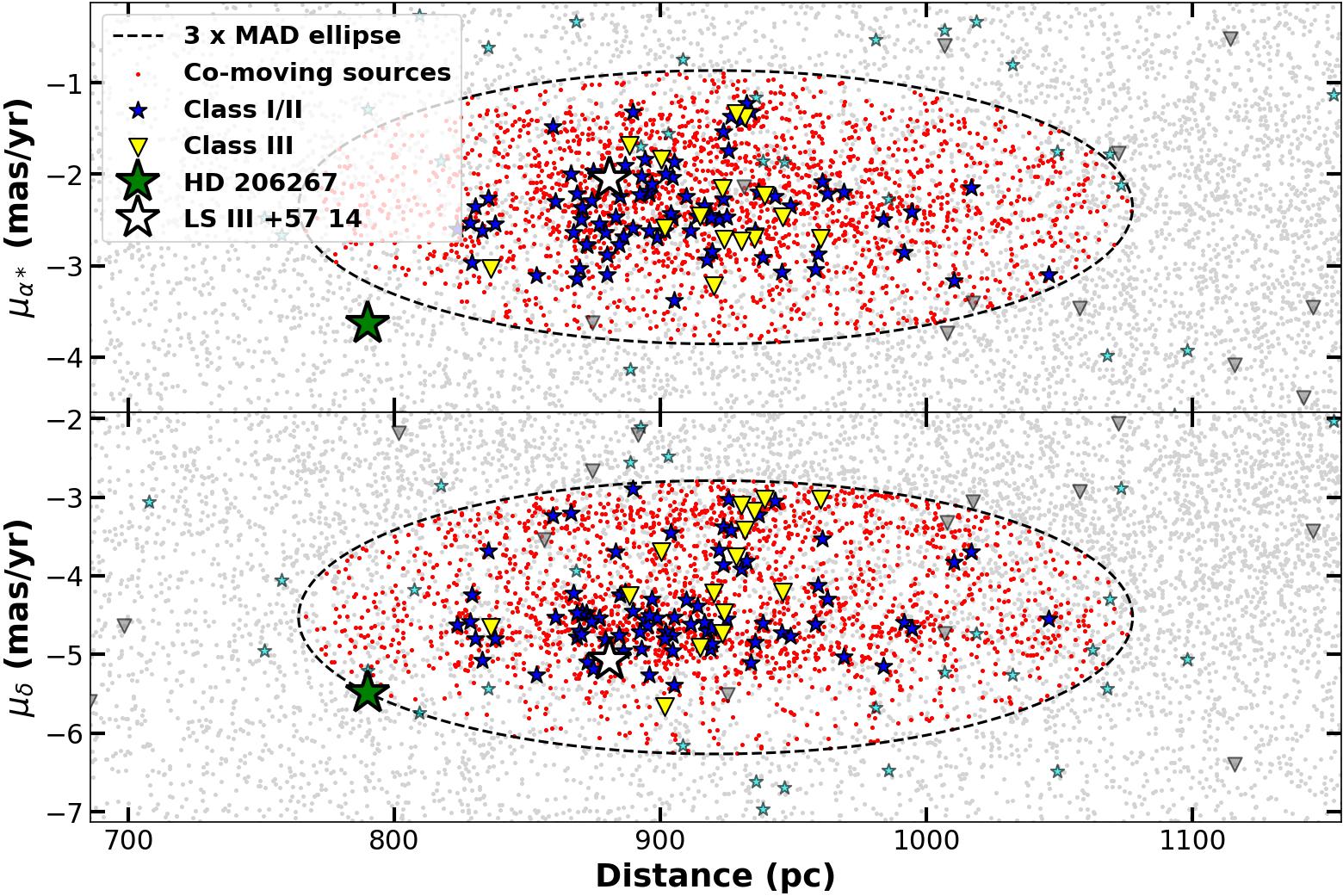}

 \caption{
    Distance versus proper motion diagrams for Collinder~69 (top) and IC~1396 (bottom).
    The dashed ellipses correspond to the WM $\pm$\,5$\times$MAD (Collinder~69) and 3$\times$MAD (IC~1396) using Class~I/II sources as reference. Co-moving candidates within the ellipses are shown as red points and the grey background sources are all sources used in the study. 
    Class~I/II YSOs inside and outside the ellipses are marked with blue and cyan stars, respectively; Class~III YSOs are shown with yellow and grey triangles. Prominent OB stars are highlighted: $\lambda$~Ori (green star) and $\lambda$~Ori~B (white star) in Collinder~69; HD~206267 (green star) and LS~III~+57~14 (white star) in IC~1396.
    }
    \label{fig:ellipse}
\end{figure}

\subsection{Cross-matching Cluster and YSO Data}

The astrometric data for cluster members from \cite{Cantat2018A&A...618A..93C} is based on {\it Gaia} DR2. To update these values to {\it Gaia} EDR3, we cross-matched the {\it Gaia} DR2 cluster data from \cite{Cantat2018A&A...618A..93C} with {\it Gaia} EDR3 data using a 1 arcsec search radius. Distance estimates were taken from \cite{bailerjohns2021AJ....161..147B}. We then selected stars with high-quality astrometry using the re-normalised unit weight error (RUWE $<$ 1.4) parameter and applied an uncertainty cut based on parallax values (parallax/error in parallax $>$ 3) \citep{Arun2021MNRAS.507..267A}.

To analyse the RE, we needed to find YSOs formed in the expanding IR bubbles surrounding the clusters, as these sources are expected to be kinematically coherent with the expansion. In both cases, the cluster radius defined by \cite{Cantat2018A&A...618A..93C} did not encompass the outer boundaries of the IR bubbles. The extended radius is estimated in the previous section. Within this estimated radius, all {\it Gaia} EDR3 sources were extracted and assessed based on the previously defined quality criteria. The newly adopted radius for Collinder 69 is 5.8\textdegree\ and for IC 1396 it is 1.4\textdegree.

The areas for both IR bubbles are large, thus we adopt the catalogue from \cite{Marton2016MNRAS.458.3479M}, which is an all sky coverage of YSOs identified from ALLWISE data. The catalogue has separate tables for class I/II and class III sources. We note that this catalogue is used here as a homogeneous all-sky YSO reference catalogue, not as a complete membership census of the two regions. Since the catalogue of \cite{Marton2016MNRAS.458.3479M} is based on WISE infrared colours, it is naturally biased towards IR-excess and disc-bearing sources, and may miss discless members and weak-excess sources, especially in distant or bright star-forming regions. Therefore, the Class~I/II sample is used only to define a young-star astrometric reference population. We extracted all the YSOs inside the updated bubble radius for both clusters. The YSOs are cross-matched with {\it Gaia} EDR3 in 1 arcsec search radius. The same astrometric quality criteria are applied to the YSOs as to the cluster members. Finally we have 151 class I/class II sources and 121 class III sources in the case of Collinder 69. In the case of IC 1396, we have 142 and 264 class I/II and class III sources, respectively.
\begin{table*}
    \centering
    \caption{Properties of the two young open clusters analysed in this study. The columns list the cluster name, equatorial coordinates, the number of member stars identified, the median proper motions with uncertainties, the adopted cluster distance with uncertainties in parsecs, the radius in degrees, and the primary ionising source responsible for the surrounding IR bubble morphology.}
    \label{tab:new}
    \begin{tabular}{lccccccccc}
        \hline
        Cluster & R.A. & Dec. & No of stars & $\mu_{\alpha *}$ & $\mu_{\delta}$ & d & Radius & Ionising source \\
        & (deg) & (deg) & & (mas yr$^{-1}$) & (mas yr$^{-1}$) & (pc) & (deg) &\\
        \hline
        Collinder 69 & 83.792 & 9.813 & 669 & $0.55 \pm 0.37$ & $-2.118 \pm 0.24$ & $402 \pm 6$ & 5.8 & $\lambda$ Ori \\
        IC 1396 & 324.745 & 57.514 & 460 & $0.398 \pm 0.46$ & $-4.495 \pm 0.38$ & $914 \pm 25$ & 1.4 & HD 206267 \\
        \hline
    \end{tabular}
\end{table*}

\begin{figure*}
    \centering
    \includegraphics[width=1.8\columnwidth]{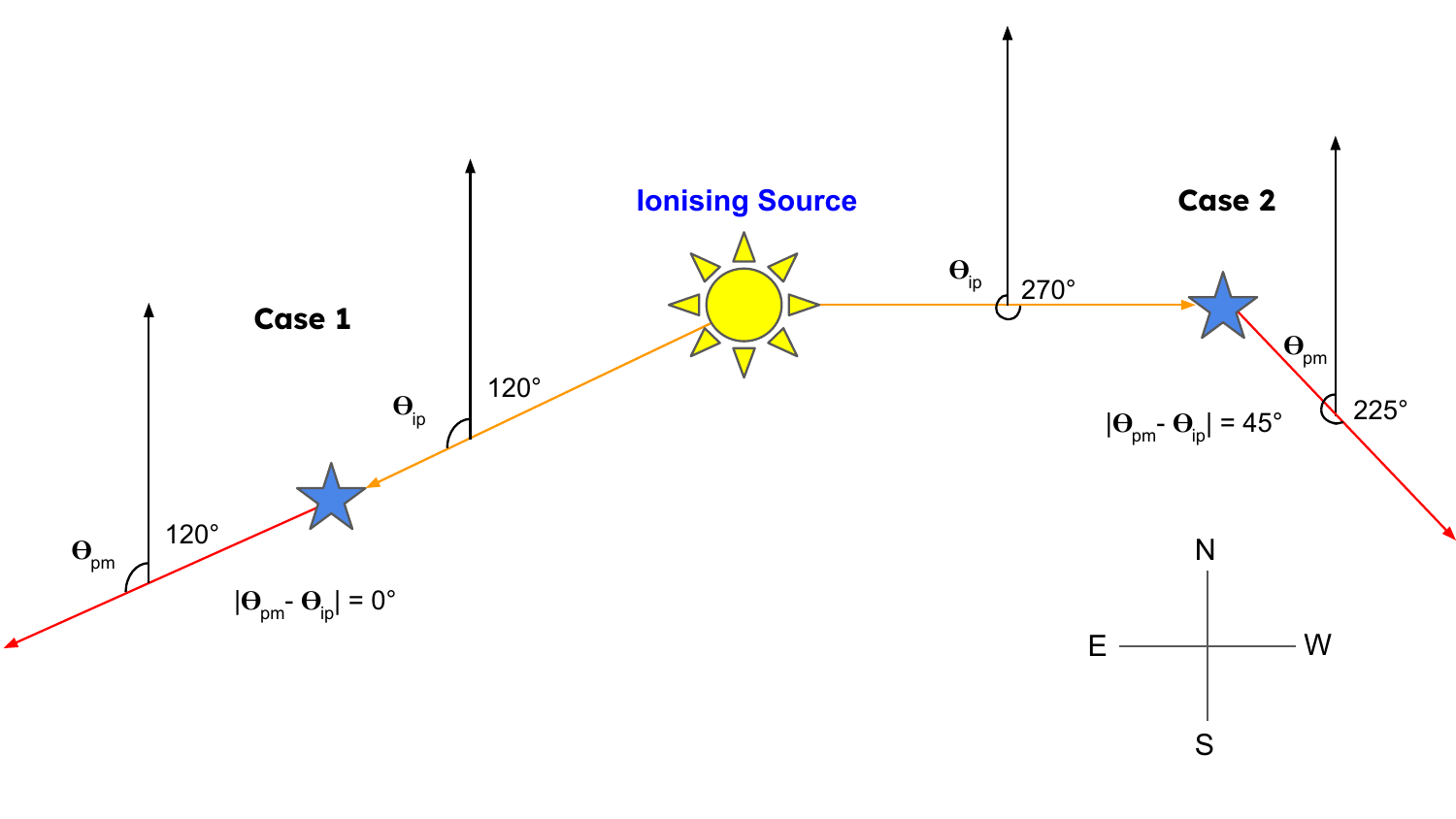}
 \caption{Schematic diagram showing a system of stars with an ionising star (yellow) at the centre. The diagram illustrates how the calculation of RPMA of stars in a cluster is performed. The angles $\theta_{ip}$ and $\theta_{pm}$ represent the angle of ionising photons and the angle of the internal relative motion of YSOs with respect to the north, respectively. The orange arrows indicate the direction of ionising photons, and the red arrows indicate the direction of the internal motion of the star. In Case 1, $\theta_{ip}$ and $\theta_{pm}$ are equal, which corresponds to the RE, as the star is moving radially away from the ionising source. Case 2 can be any other random angle from 0$-$180\textdegree. }
    \label{fig:rock}
\end{figure*}

\subsection{Membership Analysis}

We found the cluster members and YSOs with good quality astrometric data for the two clusters. In previous studies by \cite{Sahaa2022MNRAS.510.2644S} and \cite{Arun2021MNRAS.507..267A}, it is noticed that there are a number of co-moving sources with similar astrometric parameters as the YSOs in star forming regions. As we want to study the complete regions in the IR bubbles, we aim to find all co-moving sources inside the IR bubbles around Collinder 69 and IC 1396. We extracted all {\it Gaia} sources inside the updated radius for both regions. The astrometric parameters of the class I/II YSOs from \cite{Marton2016MNRAS.458.3479M} were used to identify potential new members extending into the immediate surroundings of the clusters. The class III sources are not used because they have very high dispersion in astrometric space in both regions compared to the cluster members.

Using the prescription from \cite{Sahaa2022MNRAS.510.2644S}, we employed the median and median absolute deviations (MAD) of astrometric parameters such as proper motions and distances to estimate the cluster parameters. By using the median and MAD values calculated from the class I/II YSOs, we identified stars with similar astrometric data in the extended region around the clusters. The astrometric ellipse selection is intended to identify a high-confidence co-moving population, rather than a complete census of all members. This approach can exclude kinematic outliers and may under-recover distinct subclusters, especially in complex regions such as IC~1396. As an external check, we compared our IC~1396 selection with the Gaia-based membership and substructure sample of \citet{Pelayo2023A&A...669A..22P}. The two selections have similar central astrometric values, although their ellipse is narrower and selects fewer sources. We therefore interpret our sample as a conservative co-moving population and not as a complete representation of all IC~1396 kinematic populations.

For Collinder 69, we used 5 $\times$ MAD of astrometric parameters to find the extended members, and 3 $\times$ MAD for IC 1396. In the case of Collinder 69, the median and MAD values of distance (d), proper motion in Right Ascension ($\mu_{\alpha *}$) and proper motion in Declination ($\mu_{\delta}$) are 394.07 $\pm$ 8.91 pc, $1.5 \pm 0.72$ mas/yr, and $-2.11 \pm 0.62$ mas/yr, respectively. The radius of the IR bubble is $\sim$ 40 pc, with respect to the median distance for Collinder 69. To get a similar depth of 40 pc we chose 5 $\times$ MAD for astrometric parameters. We found 2544 co-moving sources that satisfy the 5 $\times$ MAD criteria for Collinder 69. They include 738 cluster members, 127 class I/II sources and 19 class III sources.  

In the case of IC 1396, d, $\mu_{\alpha *}$ and $\mu_{\delta}$ are 920.72 $\pm$ 52.23 pc, $-2.36 \pm 0.50$ mas/yr, and $-4.53 \pm 0.58$ mas/yr, respectively. Also, the radius of the IR bubble is $\sim$ 25 pc, with respect to the median distance for IC 1396. With the increase in distance the dispersion in astrometry also increases, thus we adopted a 3 $\times$ MAD for the parameters. We found 2150 co-moving sources that satisfy the 3 $\times$ MAD criteria for IC 1396. They include 1196 cluster members, 82 class I/II sources and 15 class III sources.

We note that the co-moving sample is not a contamination-free membership list. It includes all known cluster members from \cite{Cantat2018A&A...618A..93C} together with the extended co-moving population selected by the astrometric ellipse, and a fraction of foreground and background stars will inevitably satisfy the same distance/proper-motion criteria and cannot be avoided. The cluster radius, as stated in the literature and the adopted IR bubble radius in this study, the total number of sources identified inside the bubble radius, the median and MAD values of both regions are provided in \autoref{tab:new}. The figures showing d vs. $\mu_{\alpha *}$ and d vs. $\mu_{\delta}$ with the MAD ellipses are given in \autoref{fig:ellipse}.

\subsection{Calculation of RPMA of stars in the open cluster}

To calculate the internal relative proper motion of stars in the cluster, we adopt the method defined by \cite{Sahaa2022MNRAS.510.2644S}. The proper motion values of individual stars, subtracted by the mean proper motion of the region, determine their internal relative motion \citep{Jones1997MmSAI..68..833J}. We can consider the proper motion of the ionising source as the proxy for the proper motion of its cluster, in our case, the O-type stars at the centre of the cluster have higher RUWE. Thus, we use the WM values of the cluster members for the calculations. \autoref{fig:rock} explains the measurement of RPMA of stars in this work. The angles $\theta_{ip}$ and $\theta_{pm}$ represent the angle of ionising photons and the angle of the internal/relative motion of YSOs with respect to the north, respectively. Both $\theta_{ip}$ and $\theta_{pm}$ are measured from north and increase eastward (anti-clockwise) as shown in \autoref{fig:rock}. We note that RPMA is not the sky position angle of the source itself, but the angular separation between the projected radial direction from the ionising source and the direction of the star's relative proper-motion vector. Two cases are explained in the figure. Case 1 is the ideal condition for the RE. The orange arrows indicate the direction of ionising photons directed radially outward. The red arrows indicate the direction of the internal motion of the star, which is produced by subtracting the proper motion of the central O-type star from the star. In Case 1, $\theta_{ip}$ and $\theta_{pm}$ are equal, thus $|$$\theta_{ip}$ -- $\theta_{pm}$$|$, the RPMA is  0\textdegree. As the star is moving in a radially outward direction, it is the case for the RE. In Case 2, RPMA = 45\textdegree, and the value can vary from 0 to 180\textdegree~for different stars based on their direction of observed proper motion. Since RPMA is defined as the smallest angular separation between the projected radial direction and the relative proper-motion direction, it naturally ranges from 0\textdegree~to 180\textdegree.

We calculated RPMA values for all stars identified in the region, using the astrometric parameters of the ionising star. For Collinder 69 and IC 1396, the central stars have high RUWE values, making them unreliable for RPMA calculation \citep{Saha2022MNRAS.515L..67S}. In these cases, we used the WM values of $\mu_{\alpha *}$, and $\mu_{\delta}$ of the cluster members.

The weights are assigned based on the magnitude of the star. The WM criterion is adopted because the uncertainty in {\it Gaia} astrometry increases with the decrease in the brightness of stars \citep{Lindegren2021A&A...649A...2L}, so a WM value relies on stars with better astrometry. Also, due to mass segregation, massive stars are located in the central regions of the cluster, making the estimates approximately represent the central astrometric value of the cluster.

With the above analysis, we have found sources that are astrometrically similar to the cluster-IR bubble system for Collinder 69, and IC 1396. We estimated the RPMA values of the stars with respect to the central source in each cluster.

\section{Results}

\subsection{RPMA distribution}

\begin{figure*}
    \includegraphics[width=\columnwidth]{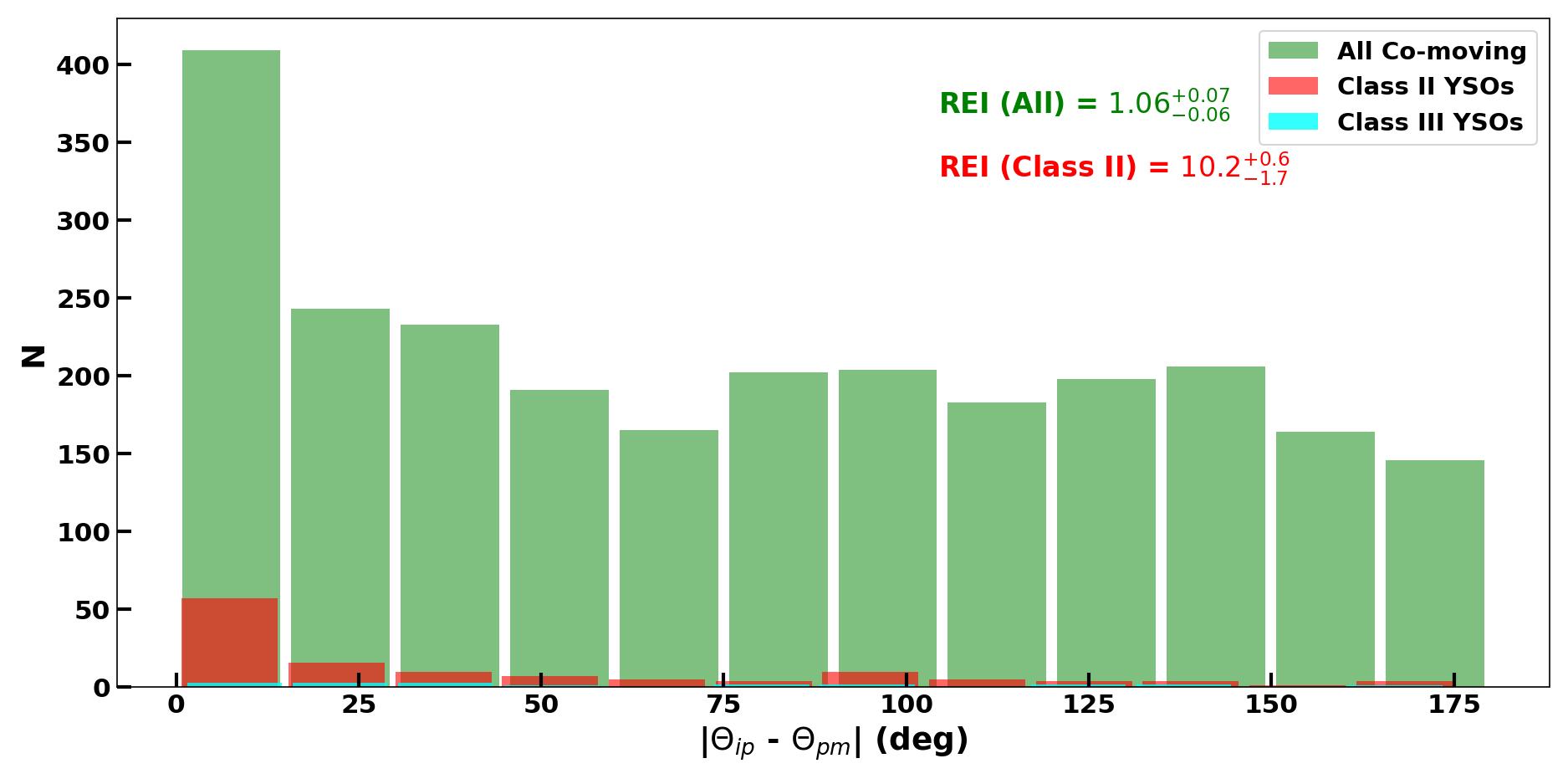}
    \includegraphics[width=\columnwidth]{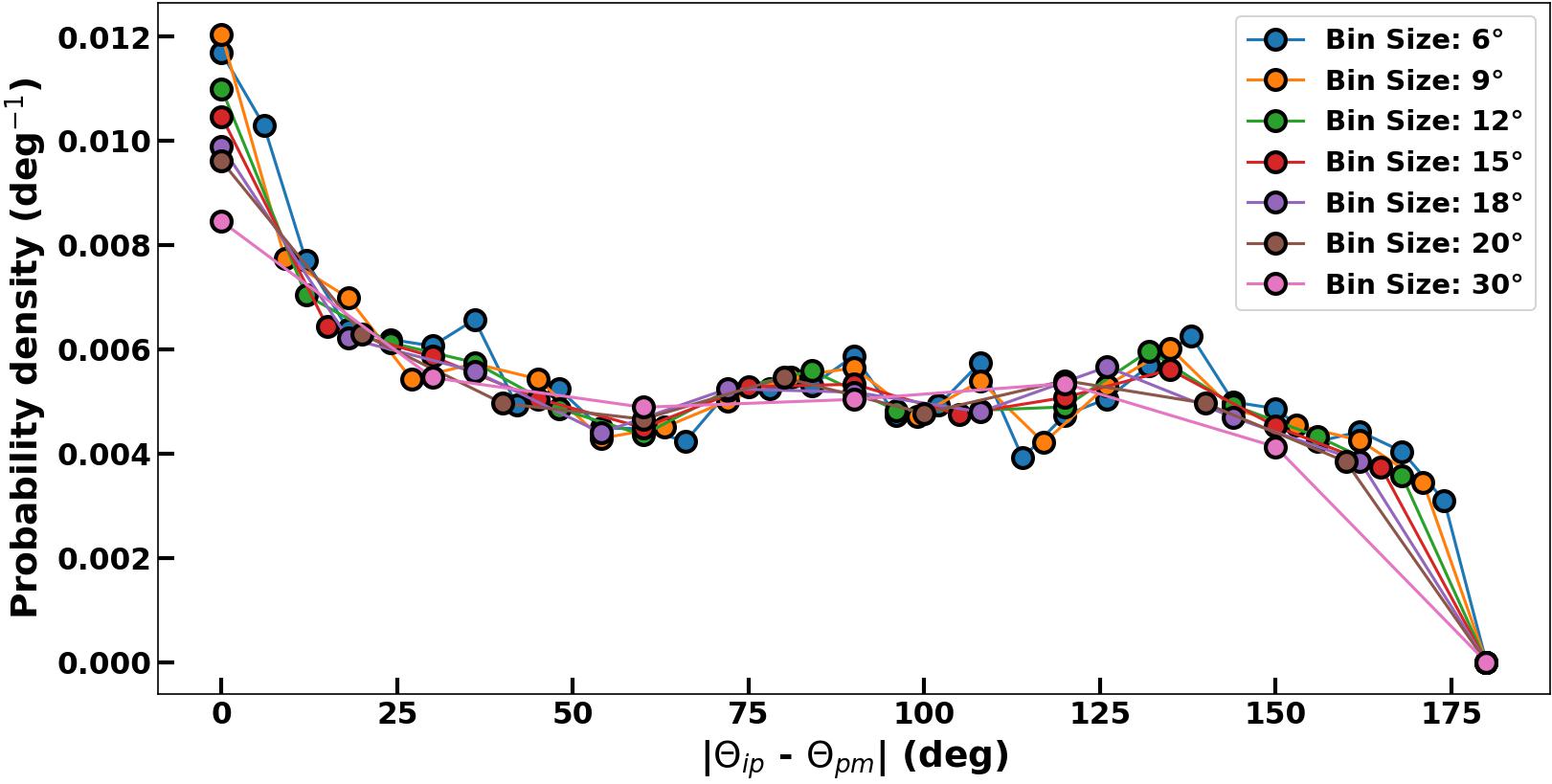}
    \includegraphics[width=\columnwidth]{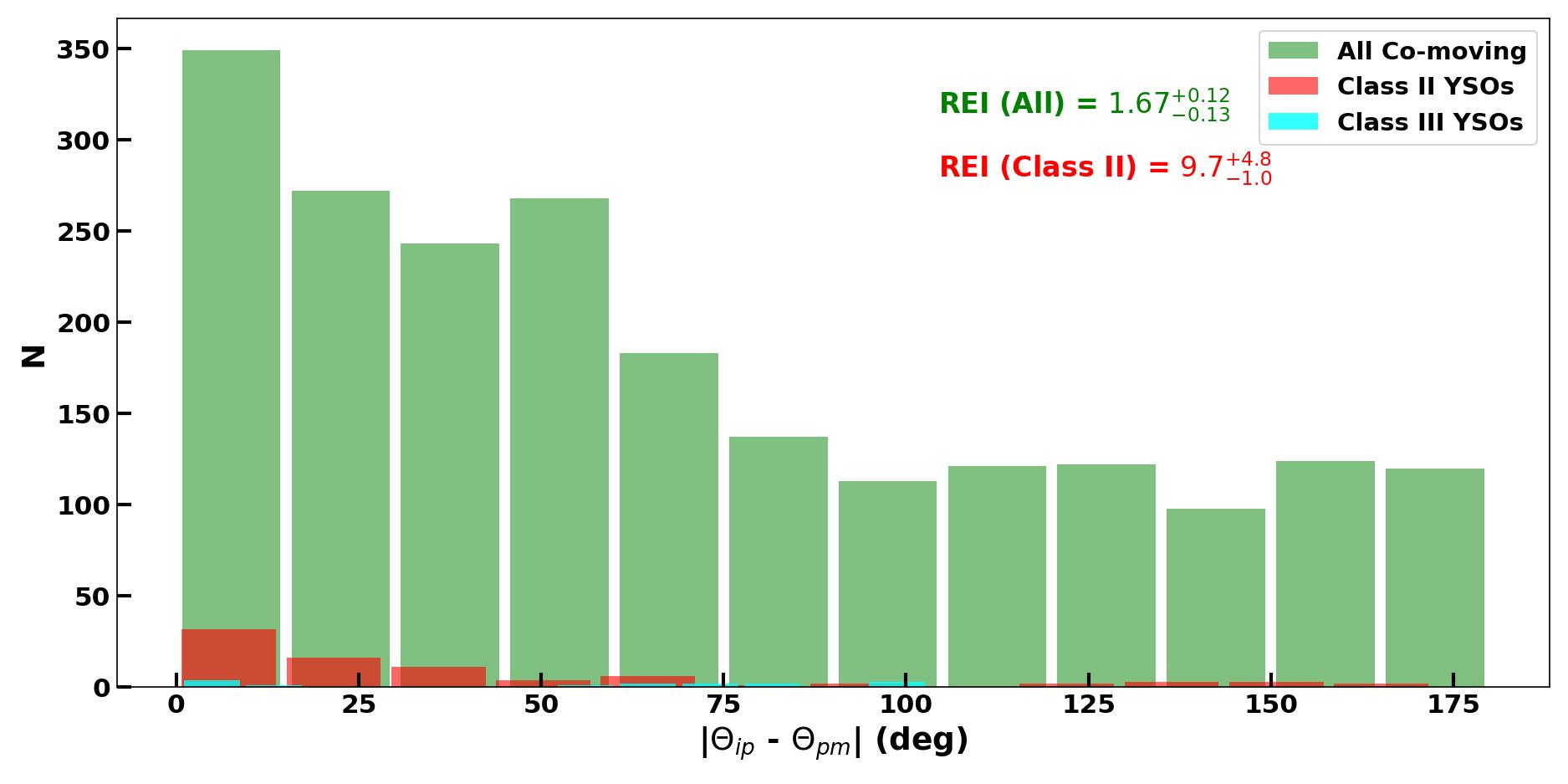}
    \includegraphics[width=\columnwidth]{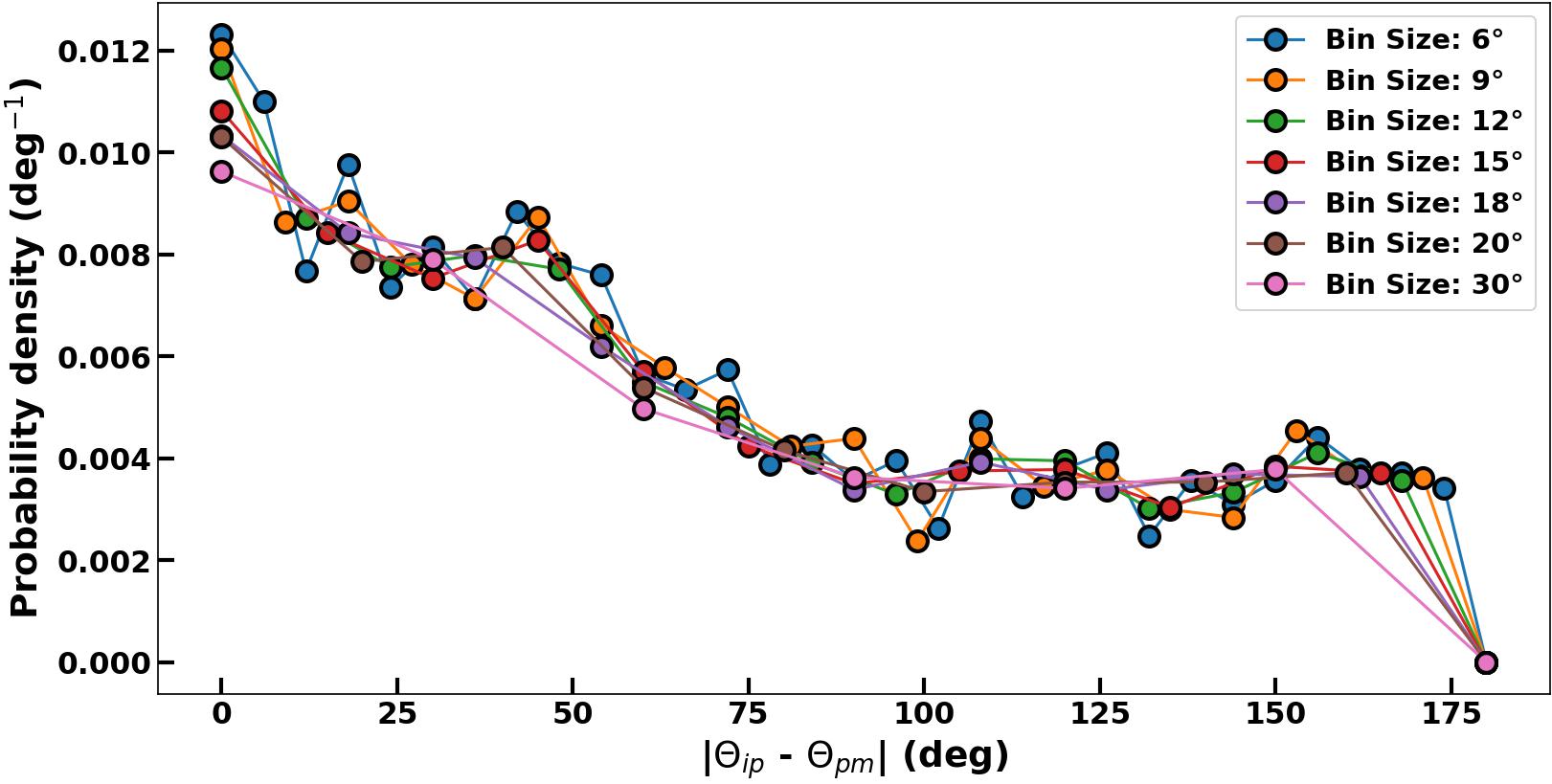}

   \caption{The figure shows the RPMA histogram distribution (left) of Collinder 69 \textbf{(top)}, and IC 1396 \textbf{(bottom)}. The distribution is divided into 12 bins, each with a 15\textdegree~interval. Green shows all co-moving sources, red the Class I/II YSOs and cyan the Class III YSOs. All histograms show that the first bin (RPMA $< 15^\circ$) is significantly higher than the median. The REI values for the full co-moving sample and for the Class I/II sources are denoted in each histogram. The positive value of REI indicates the presence of a radially outward-moving population within these young clusters. Additionally, the subsequent bins in each histogram show a gradual decrease in the number of stars, eventually plateauing. The top and bottom panels on the right show the bin sensitivity analysis for Collinder 69 and IC 1396 respectively, giving the probability density of RPMA (deg$^{-1}$) for bin widths from 6\textdegree~to 30\textdegree}, which shows the result does not deviate with different bin sizes.
    \label{fig:PPMAHIST}
\end{figure*}

RPMA is the kinematic angle measuring the alignment between the projected radial direction from the ionising source and the star's relative proper-motion vector, and not the angular extent of the cluster or the sky position angle of a source. We calculated the RPMAs which span from 0 $-$ 180 \textdegree\ for all stars that are found to be associated with both clusters. The 0\textdegree\ angle means the stars are moving radially away from the ionising source. The 180\textdegree\ angle corresponds to motion in the opposite direction, i.e. towards the ionising source. The histogram distribution of the angles for both clusters is shown in \autoref{fig:PPMAHIST}. The histograms have 12 bins, each representing a 15\textdegree~interval. We observe an elevated number of stars in the 0$-$15\textdegree~bins in both clusters, indicating a significant fraction of sources moving away from the central source. If the relative proper-motion directions were randomly oriented with respect to the radial direction from the ionising source, the RPMA distribution would be approximately uniform across the 0--180\textdegree~range. Assuming a random distribution of angles for stars in a cluster, the histogram distribution should be approximately equal across all bins. For example, in the case of Collinder 69, the median frequency of all bins excluding the first bin is 198. The first bin, which contains 407 stars, is approximately 207\% of that median. Similarly, for IC 1396, the median excluding the first bin is 124, while the first bin contains 349 stars, which is approximately 281\% of the median. This clearly indicates an over-density of sources moving radially away (RPMA $< 15^\circ$) from the centre in both clusters.

To quantify the directional nature of the deviation, we define and estimate the Rocket Effect Index (REI). The REI is designed to measure the prominence of outward-moving stars by comparing the number of stars in the first angular bin (0--15\textdegree) to the median frequency of stars in the other 11 bins. The first bin is chosen because RPMA $<15\textdegree$ represents the most strongly outward-moving population expected from the rocket effect, while the median of the remaining bins provides a robust estimate of the typical non-preferred angular population. The REI is defined as:

\begin{equation}
\text{REI} = \frac{\text{No. of stars in first bin} - \text{Median frequency of other 11 bins}}{\text{Median frequency of other 11 bins}}.
\end{equation}

To assess the effect of proper-motion uncertainties on the REI values, we performed a Monte Carlo propagation of the Gaia proper-motion errors. In each realisation, the individual $\mu_{\alpha *}$ and $\mu_{\delta}$ values were perturbed within their quoted uncertainties, the RPMA values were recalculated, and the REI was recomputed. This was repeated 1000 times, and the median and 16th--84th percentile range of the resulting REI distribution were adopted as the uncertainty on REI. For instance, in the case of Collinder 69, the REI is calculated as $\bm{1.06^{+0.07}_{-0.06}}$, indicating that the first bin contains $\sim$ 100\% more stars than the expected number of stars. Similarly, for IC~1396, the REI is $\bm{1.67^{+0.12}_{-0.13}}$. We also estimated RPMA values for the Class~I/II sources in our sample. It is interesting to note that the REI values for Class~I/II sources alone are $\bm{10.2^{+0.6}_{-1.7}}$ and $\bm{9.7^{+4.8}_{-1.0}}$, respectively, for Collinder~69 and IC~1396. This is indicative of triggered star formation through RE as disc-bearing sources are predominantly moving away from the ionising O-type stars.

We also conducted a sensitivity analysis to evaluate the impact of bin size on the RPMA histogram distributions and ensure the robustness of our findings. By varying the bin size across 6\textdegree, 9\textdegree, 12\textdegree, 15\textdegree, 18\textdegree, 20\textdegree, and 30\textdegree, the density of stars per bin was calculated as the frequency of stars within each angular interval normalised by the bin width. Specifically, for each bin size (e.g., 6\textdegree, 9\textdegree, 12\textdegree, etc.), the number of stars falling within that angular range was counted and divided by the bin width to ensure comparability across different bin sizes. This density provides a consistent way to compare the relative over-densities in the first bin (RPMA $<$ 15\textdegree) across varying resolutions. IC 1396 and Collinder 69 consistently exhibited a peak in the first bin, regardless of the bin size. This stability underscores the robustness of the observed over-density of stars with RPMAs $<$ 15\textdegree.

Additionally, it is noteworthy that the RPMA distribution of both clusters shows that the frequency in each bin gradually decreases and normalises by the fifth bin (60 - 75\textdegree). This also indicates the presence of a radially outward-moving population within these young clusters.

To confirm that the RPMA excess in the first bin is significant independently of the REI definition, we applied three further statistical tests. First, considering under an isotropic null, the probability that a source falls in the first 15\textdegree~bin is $p_0 = 15\textdegree/180\textdegree = 1/12$. A one-sided binomial test on the observed first-bin counts (407/2544 for Collinder~69, 349/2150 for IC~1396, i.e. 1.9$\times$ the isotropic expectation in both) rejects isotropy at $p = 2.4\times10^{-36}$ and $1.2\times10^{-32}$, respectively. Second, we tested the RPMA distribution for departures from a Uniform$[0,180\textdegree]$ distribution using an Anderson--Darling test \citep{AndersonDarling1952}, which is sensitive to excesses in the low-RPMA tail; this yields $A^2 = 72.9$ (Collinder~69) and $A^2 = 179.9$ (IC~1396), far above the 1\% critical value of 3.86. Third, because the rocket effect specifies the expected direction a priori (radially outward), we applied a V-test, a variation of the Rayleigh test against a specified mean direction, to the signed offset angle \citep{GreenwoodDurand1955,Batschelet1981}. This is the most appropriate directional statistic when the target direction is already known. The V-test gives $V = 0.137$ ($u = 9.8$, $p = 8\times10^{-23}$) for Collinder~69 and $V = 0.278$ ($u = 18.2$, $p = 2\times10^{-74}$) for IC~1396, with mean relative proper-motion directions of 3.9\textdegree~and 8.3\textdegree~from the outward radial. Together, these tests establish that the outward-motion excess is statistically significant and directionally aligned with the ionising source, independently of the REI. In the case of the Class~I/II population in both regions, the directional signal is stronger. The V-test amplitude increases to $V = 0.58$ for Collinder~69 and $V = 0.65$ for IC~1396, about two to four times the values for the full co-moving samples, while the Anderson--Darling test remains a decisive rejection of uniformity ($A^2 = 72.5$ and $A^2 = 53.2$, respectively). This is indicative of the disc-bearing YSOs being more coherently aligned with the outward radial direction than the general co-moving population.

\begin{figure}
    \includegraphics[width=\columnwidth]{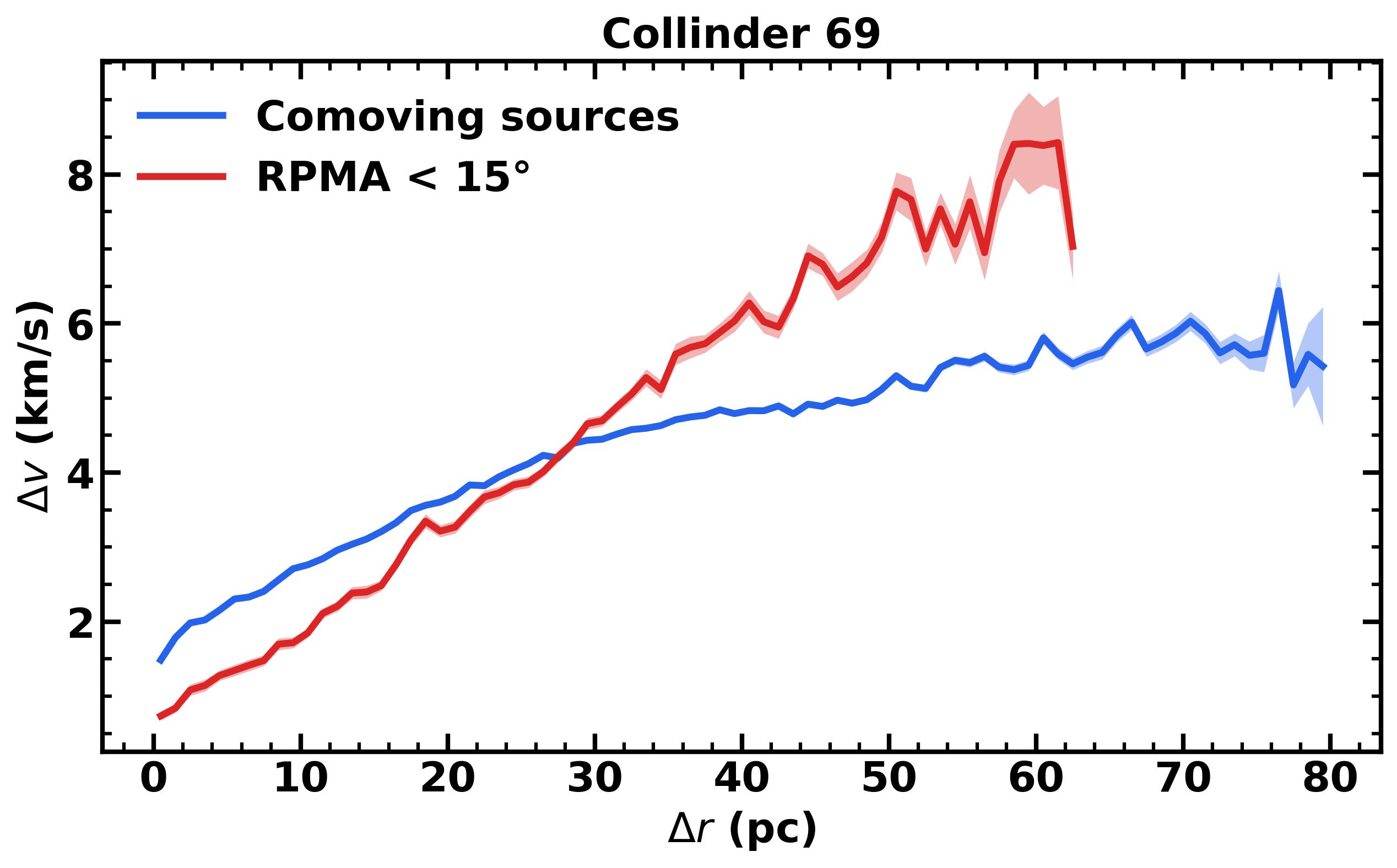}
    \includegraphics[width=\columnwidth]{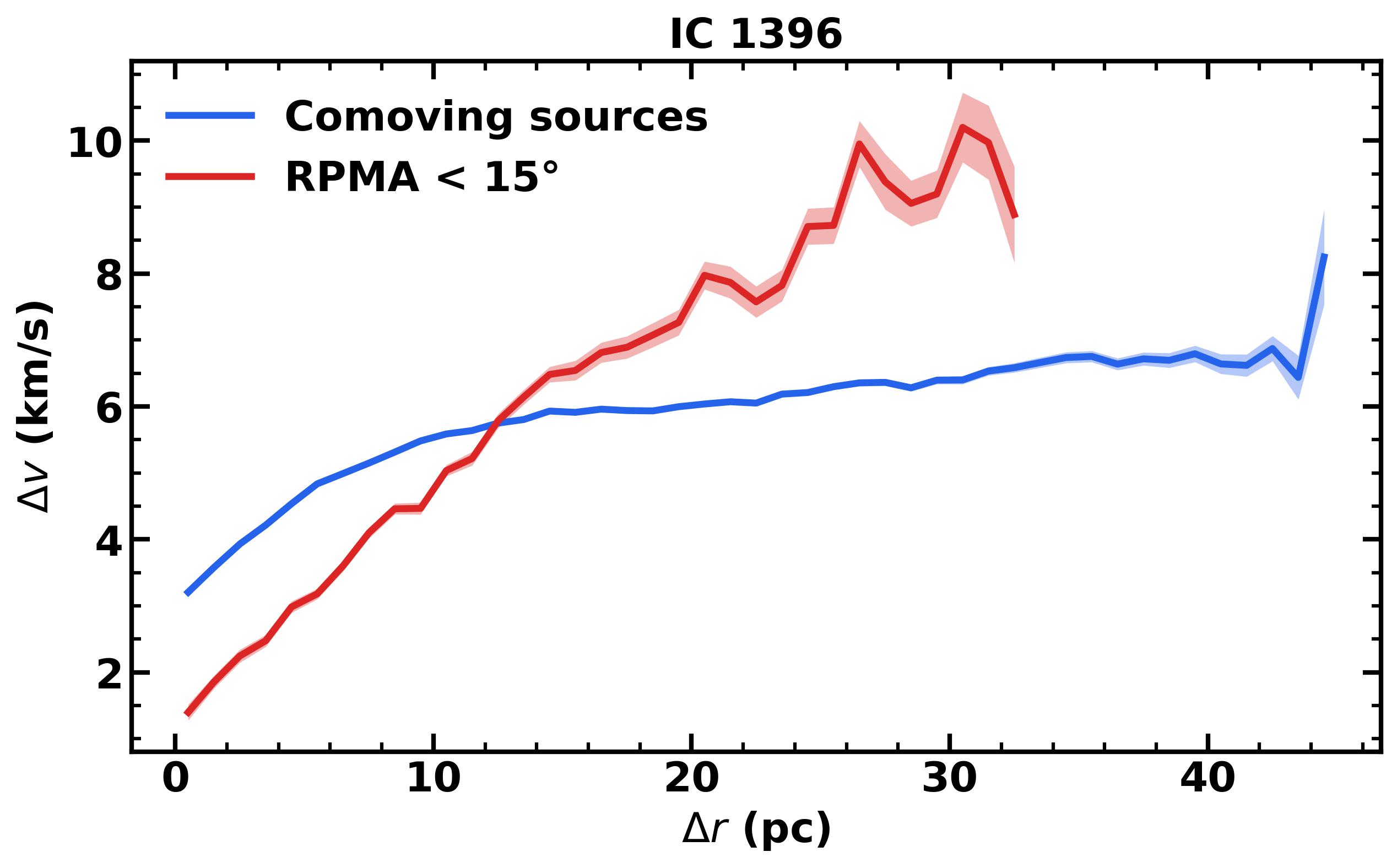}
    
    \caption{Velocity structure analysis for Collinder 69 and IC 1396. The mean pairwise velocity difference, $\Delta v$, is shown as a function of pairwise spatial separation, $\Delta r$. Blue curves show the full comoving-source samples, while red curves show the subsamples with $\mathrm{RPMA}<15^\circ$. Shaded regions indicate the $\pm3\sigma$ uncertainty band on the mean $\Delta v$ in each separation bin.}   
    \label{fig:vsat}
\end{figure}

To compare the kinematic properties of the two young clusters with a broader open-cluster reference sample, we used the catalogue by \cite{Hunt2023A&A...673A.114H}, restricted to clusters that are also present in the \cite{Cantat2018A&A...618A..93C} catalogue used in our initial target selection. \cite{Cantat2018A&A...618A..93C} provides the original Gaia DR2 open-cluster basis for our ROCKETS selection, while the catalogue of \cite{Hunt2023A&A...673A.114H}  provides updated Gaia EDR3-based membership lists and structural parameters. In particular, the \cite{Hunt2023A&A...673A.114H} catalogue provides core radii, which are required here to define the central reference astrometry through the median position and proper motion of core members. We selected common clusters with at least 100 members and distances below 1.5~kpc, yielding 314 clusters for which the RPMA calculation could be performed. For each reference cluster, stars inside the catalogue core radius were used only to define the central astrometric reference, and the RPMA and REI were then computed using the cluster members.

As shown in the Kernel Density Estimate (KDE) plot of REI (\\autoref{fig:kde} of Appendix~\ref{app:kde}), the external open-cluster reference sample exhibits distributions centred around low positive values for REI, with a median REI of 0.150. These values provide a reference level for the low-RPMA excess expected in nearby Gaia open clusters, but this sample is not a physically matched control sample of young O-star-bearing IR bubbles. Additionally, the majority of clusters in the reference sample show low-RPMA fractions close to the isotropic expectation of $1/12$, indicating little preferential outward motion.

In contrast, Collinder~69 and IC~1396 exhibit higher REI values than the median of the external reference sample. The REI values indicate a pronounced over-density of stars with RPMAs in the 0--15\textdegree~range. The low-RPMA fraction is used only as a descriptive comparison metric and not as evidence by itself for the physical origin of the signal. Our results align closely with the findings of \cite{Kuhn2019ApJ...870...32K} and \cite{Armstrong2024A&A...692A.166A}. \cite{Kuhn2019ApJ...870...32K} demonstrated that a majority of young clusters exhibit significant expansion, driven by gas expulsion following feedback from massive stars. These results and previous studies consistently indicate a kinematic signature of radially outward motion in the young clusters, especially by the disc bearing sources, which could be due the feedback driven process from massive O-type stars.

\begin{figure*}
    \includegraphics[width=\columnwidth]{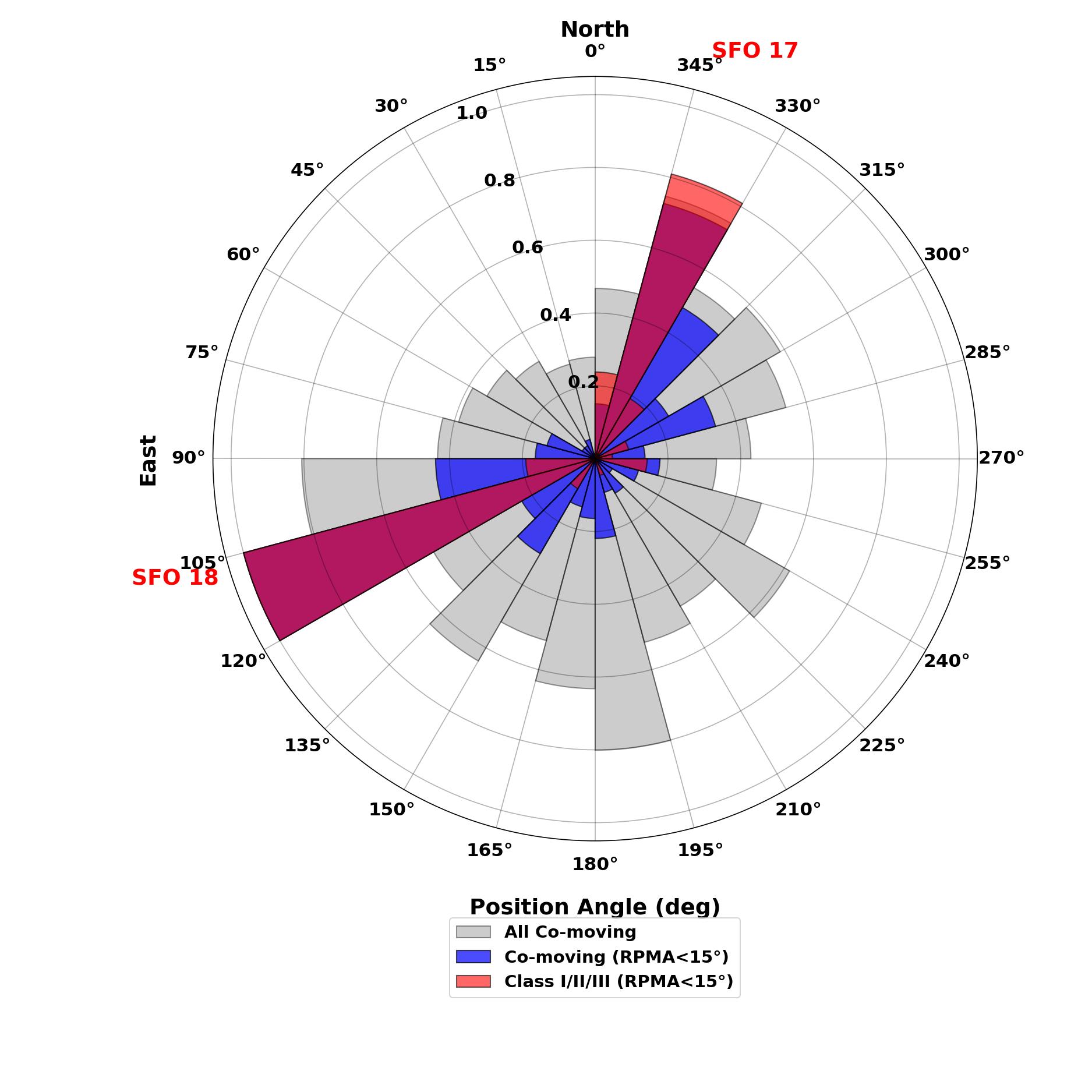}
    \includegraphics[width=\columnwidth]{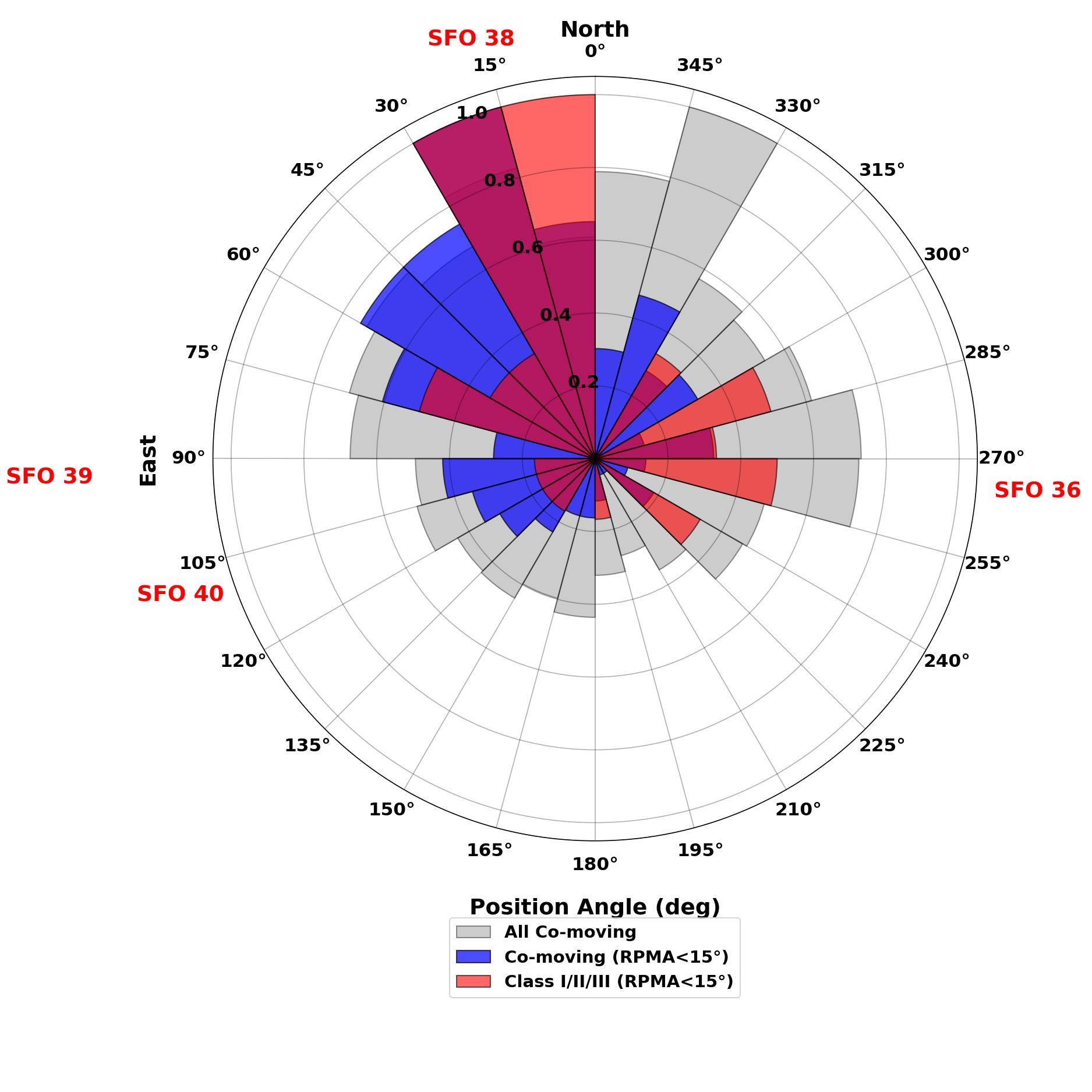}
    
    \caption{The figure shows the PA polar histogram distributions of cluster members from the cluster centre of Collinder 69, and IC 1396 respectively. The PA histogram show that most of the RPMA $< 15^\circ$~ stars are predominantly directed towards known BRCs and not isotropically distributed. Unlike RPMA, the PA shown here represents the sky position angle of the sources with respect to the cluster centre. The histograms show all sources satisfying astrometric ellipse (grey), all RPMA $< 15^\circ$~sources (blue) and Class I/II/III sources with RPMA $< 15^\circ$~. For visualisation the histograms are normalised to 1 on the highest bin. }
    \label{fig:polar}
\end{figure*}

\subsection{Comparison with VSAT}

To compare the results from previous section with an independent velocity-structure diagnostic, we performed a Velocity Structure Analysis Tool (VSAT) analysis following \citet{Arnold2019MNRAS.483.3894A}. VSAT quantifies the velocity structure of a stellar population by considering all possible pairs of stars. For each pair, the projected separation $\Delta r$ and velocity difference $\Delta v$ are calculated. The method defines two velocity statistics: the magnitude velocity difference, $\Delta v_M$, and the directional velocity difference, $\Delta v_D$. The latter measures the rate at which the separation between a pair of stars is changing. Positive $\Delta v_D$ indicates that the pair is moving apart, while negative $\Delta v_D$ indicates that the pair is moving towards each other.

For Collinder~69 and IC 1396, we applied this analysis to the full co-moving population used in the RPMA analysis and separately to the subset of stars with RPMA $<15^\circ$. The sky positions were converted into projected physical positions relative to the adopted cluster centre, and the proper motions were converted into tangential velocities using the Bailer-Jones geometric distances. We then computed $\Delta r$, $\Delta v_M$, and $\Delta v_D$ for all stellar pairs and binned the resulting values as a function of projected separation.

The comparison is shown in \autoref{fig:vsat}. The full co-moving population shows mostly positive $\langle \Delta v_D \rangle$ over a large range of projected separations, indicating an overall expansion-like velocity structure. More importantly, the RPMA $<15^\circ$ subset shows systematically higher positive $\langle \Delta v_D \rangle$ than the full co-moving sample across most separations. This indicates that the stars selected as radially outward-moving with respect to the ionising source are also more strongly expanding in the pairwise VSAT sense.

VSAT alone does not identify the rocket effect, because it measures pairwise expansion or contraction and is not tied to a specific ionising source. In contrast, the REI from previous section is explicitly referenced to the projected radial direction from the central O-type star. The fact that the RPMA $<15^\circ$ subset also shows stronger pairwise expansion supports the interpretation that the REI-selected population traces a genuine outward-moving component associated with feedback from the ionising source.

\subsection{Position Angle Distribution of RPMA stars}

Both clusters, with feedback from O-type stars, show a significantly higher number of stars moving radially away (RPMA $< 15^\circ$) from the cluster centre. However, it would be an over-interpretation of the data to assume that all stars with a RPMA $< 15^\circ$~are influenced by the RE. It is more realistic to assume that cluster members also show random motion in the 2-dimensional plane that we observe. Thus, a percentage of sources in the first bin (RPMA $< 15^\circ$) could be random components from the cluster members. It is difficult to remove these random components from our analysis with the present data. Also, \cite{Armstrong2024A&A...692A.166A}, who investigated the expansion of the Collinder 69 ($\lambda$ Ori) cluster found that the expansion of $\lambda$ Ori is anisotropic, with stellar motions showing a clear directional dependence rather than isotropy. \cite{Armstrong2024A&A...692A.166A} attribute this anisotropy to the feedback-driven expulsion of gas, which accelerates stars preferentially along certain directions. Therefore, to analyse if anisotropy can be found in our study, we created polar histogram distributions of the position angle of stars with respect to the cluster centre. The histograms are normalised to the highest frequency bin for visual purposes.

\subsubsection{Polar Histograms}

\autoref{fig:polar} shows the polar histograms of Collinder 69 and IC 1396. The polar histograms display all stars used in the study (pink) and stars with RPMA $< 15^\circ$~(blue) separately. In the case of Collinder 69, the stars with RPMA $< 15^\circ$~are not uniformly distributed in all directions but are highly inclined towards the East and West, confirming the anisotropy found by \cite{Armstrong2024A&A...692A.166A}. In the eastern direction, most stars are directed towards BRC 18. Similarly, a significant number of sources are directed towards BRC 17. The populations directed towards these BRCs have been individually studied and established by \cite{Sahaa2022MNRAS.510.2644S} and \cite{Saha2022MNRAS.515L..67S}. The polar histogram also shows this directional bias towards the BRCs. The position angles of SFO 17 and 18 are 105 and 338.5\textdegree~respectively. We observe a high number of sources, approximately 32\%, directed towards SFO 18 and 9\% of stars directed towards SFO 17. The overall distribution of stars also shows a high number of stars towards the BRCs.

For IC 1396, the sources with RPMA $< 15^\circ$~are found in all directions of the cluster. Previous studies have also reported non-random distributions of young stars associated with clouds, globules, and ionisation fronts in IC~1396 \citep{Reach2004ApJS..154..385R,Sicilia-Aguilar2006AJ....132.2135S}. Our result is consistent with these studies, but adds the Gaia-based kinematic evidence for preferential outward motion. In the present analysis, there is a clear over-density of sources pointing towards BRCs, SFO 36 (Elephant Trunk) to the west, and SFO 38 to the north. These BRCs are also studied individually by \cite{Saha2022MNRAS.515L..67S}. The position angles of SFO 36, 38, 39, and 40 are 261, 17.9, 93.7, and 109.5\textdegree~respectively. The polar histogram also shows a high number of sources angled towards these BRCs, proving anisotropy.

\subsubsection{Relative Proper Motion Vectors}

\begin{figure*}
    \includegraphics[width=\columnwidth]{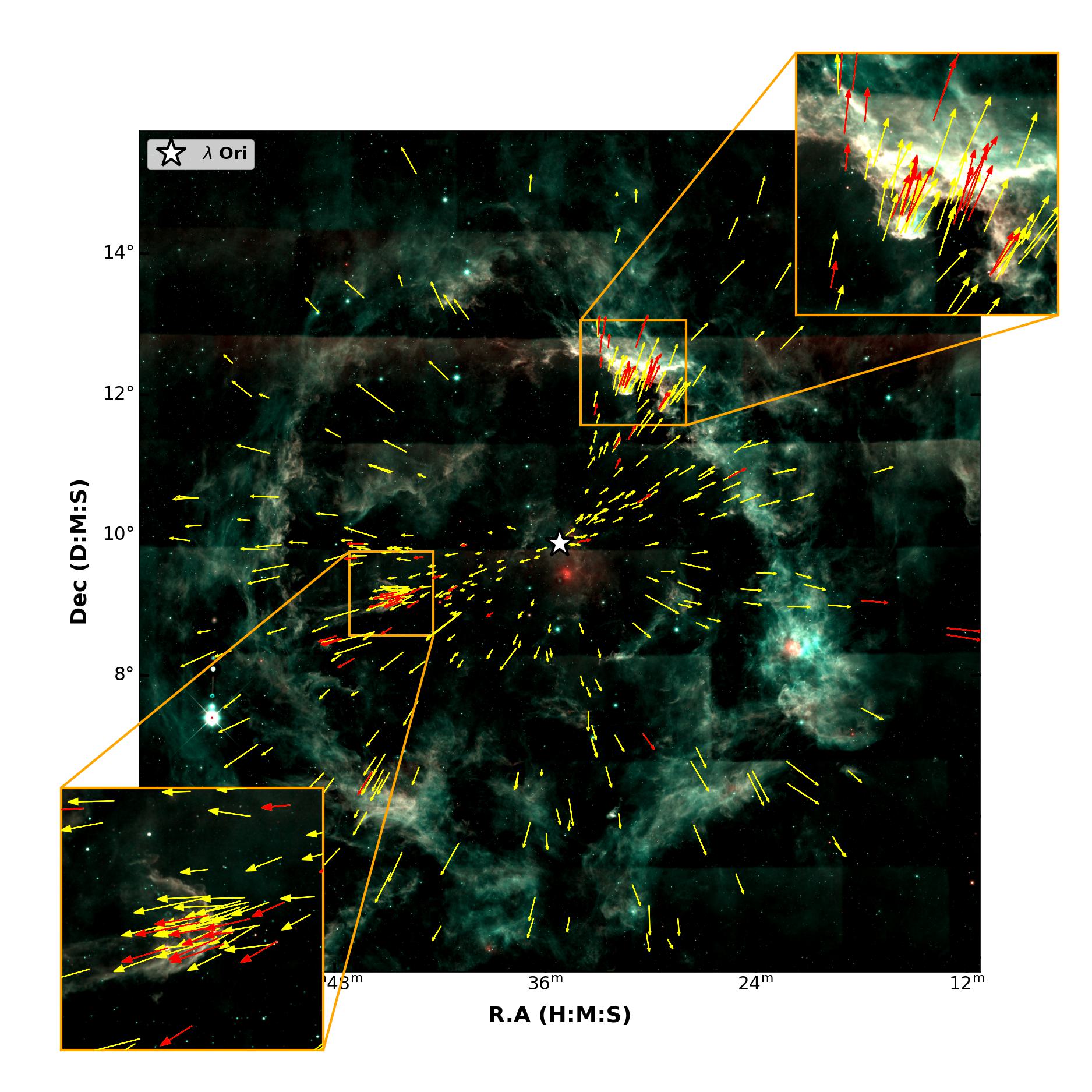}
    \includegraphics[width=\columnwidth]{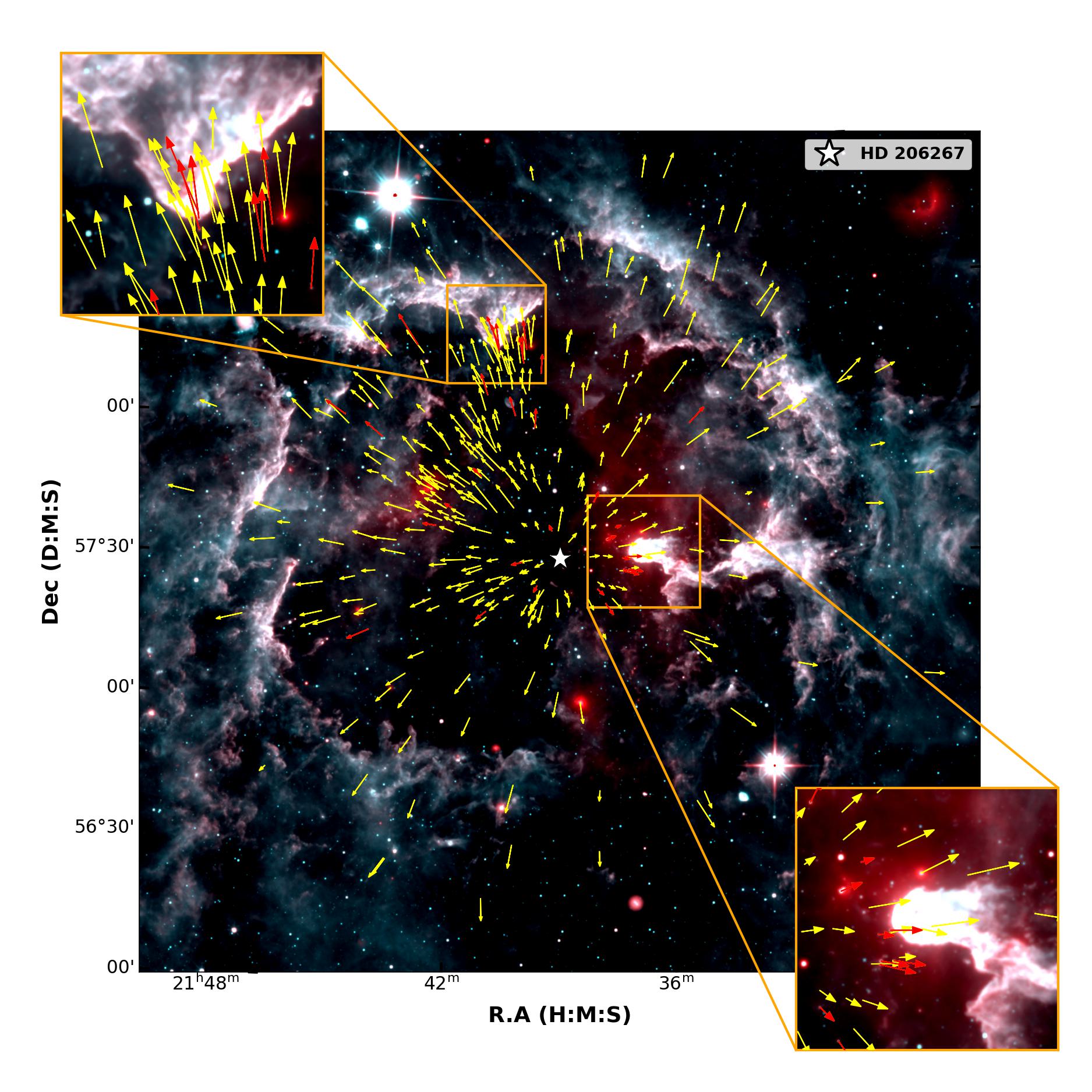}
    
    \caption{WISE colour–composite (12$\mu m$ W3 band in turquoise and 24 $\mu m$ W4 band in red) images of Collinder 69 and IC 1396 overlaid with stellar relative proper–motion vectors, measured with respect to the weighted median proper motion of the cluster members. Yellow arrows indicate all sources with RPMA $<15^{\circ}$, while red arrows mark Class I/II/III YSOs with RPMA $<15^{\circ}$. Orange boxes show zoomed views towards the BRCs.}
    \label{fig:color}
\end{figure*}

The same analysis is visualised on the colour–composite images of the Collinder 69 and IC 1396 regions overlaid with stellar relative proper–motion vectors (\autoref{fig:color}). The background image is constructed from WISE bands (12$\mu m$ W3 band in turquoise and 24 $\mu m$ W4 band in red), highlighting the IR bubble boundaries and BRCs. Relative proper motions are shown as arrows, with yellow arrows indicating all sources with RPMA $< 15^{\circ}$ and red arrows representing Class I/II/III YSOs with RPMA $< 15^{\circ}$. To emphasise BRCs with enhanced source density, two inset panels (orange boxes) provide zoomed–in views, revealing clear over–densities of sources towards the BRCs in both clusters.

Our analysis shows that stars with RPMA $< 15^\circ$~have preferential distribution in both clusters and are not uniformly distributed (anisotropic). We find that a major share of stars are directed towards known BRCs and bright rim sources and expanded away regions, proving RE and gas expulsion through feedback as suggested by \cite{Armstrong2024A&A...692A.166A}. This suggests that the RE does not uniformly act in all directions but rather in specific directions during the younger stages of star formation. The acceleration process/shock fronts and density variation forms BRC structures towards particular directions. \cite{Dale2015} argued that the stars formed in these accelerating clouds exhibit radially outward motion, consistent with the direction of the clouds' acceleration. Interestingly, we also observe sources with radially outward motion between the BRCs and the ionising star, indicating that the stars formed in these cloud trails are also being accelerated away. The grouping of these sources near the globules is consistent with the distinct kinematic populations and successive star-forming episodes reported in IC 1396 by \citet{Pelayo2023A&A...669A..22P}, and by \citet{Sanju2024MNRAS.534.2566S} for Collinder 69.

\subsubsection{Radial Distribution of Sources}

\begin{figure}
    \includegraphics[width=\columnwidth]{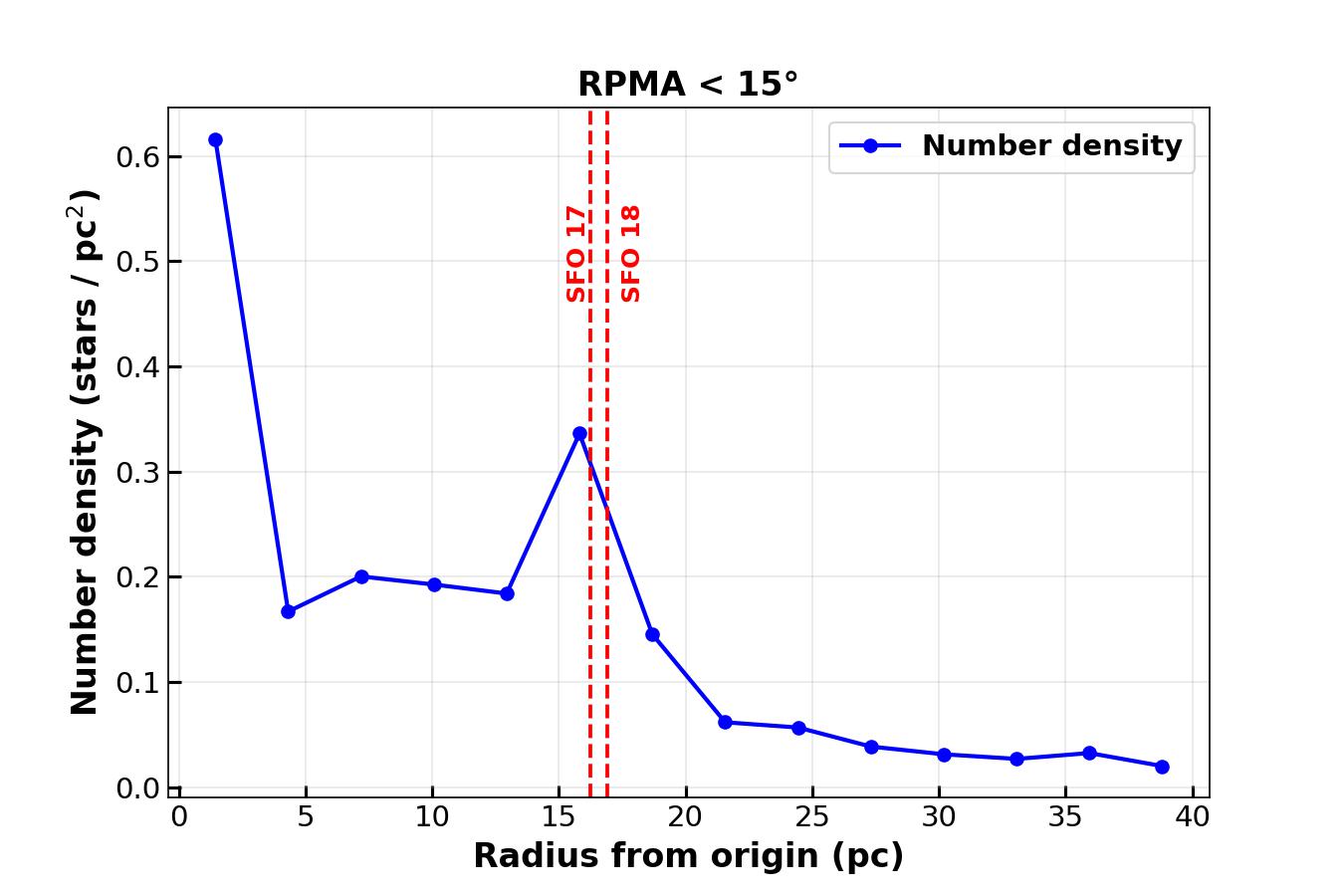}
    \includegraphics[width=\columnwidth]{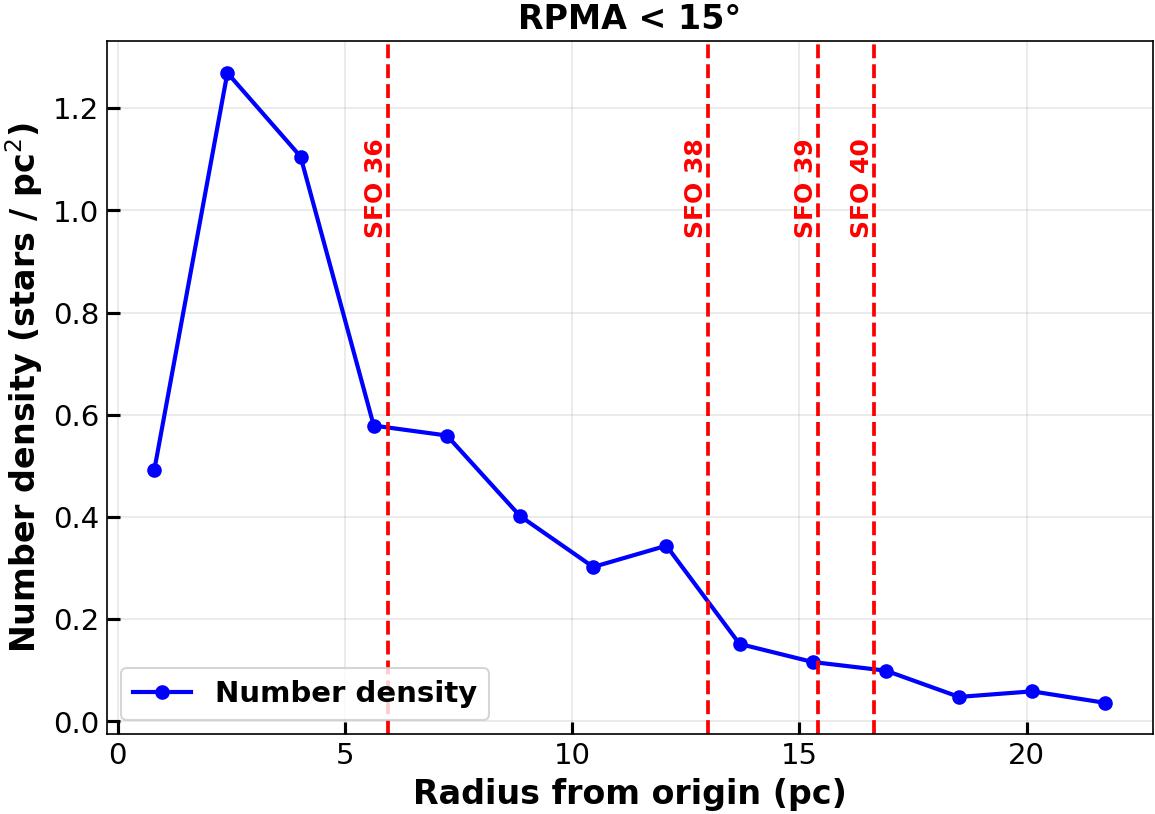}

\caption{
        Radial number density profiles of stars with RPMA $<$ 15\textdegree. 
        \textbf{Top:} Collinder~69, showing an enhancement at $\sim$15\,pc that coincides with SFO~17 and SFO~18. 
        \textbf{Bottom:} IC~1396, where elevated densities are seen at $\sim$6\,pc and $\sim$13\,pc corresponding to SFO~36 and SFO~38. 
        In both clusters, the overall decline towards the outskirts indicates that the adopted radii encompass the main stellar populations, while localised density enhancements near bright-rimmed clouds provide additional evidence for the RE.
    }
    \label{fig:numdens}
\end{figure}

To further examine the spatial distribution of stars associated with the RE, we computed the radial number density profiles of sources with RPMA $<$ 15\textdegree~relative to the cluster centres (see \autoref{fig:numdens}). In both clusters, the number density declines smoothly towards the outskirts, indicating that our adopted cluster radii are sufficient and that no significant population was missed beyond these boundaries.

In Collinder~69, the profile shows a clear enhancement in stellar number density at a projected distance of $\sim$15 pc from the ionising source. This coincides with the locations of SFO~17 and SFO~18, which were already shown in the previous subsection to host highly directional stellar density. Similarly, in IC~1396, elevated stellar densities are found at projected distances of $\sim$6 pc and $\sim$13 pc, corresponding to BRCs SFO~36 and SFO~38. The alignment of these density enhancements with known BRCs provides further support for the RE as the driver of both stellar kinematics and clustered star formation in these regions.

\subsection{Colour-Magnitude Diagram Analysis}

\begin{figure*}
    \includegraphics[width=\columnwidth]{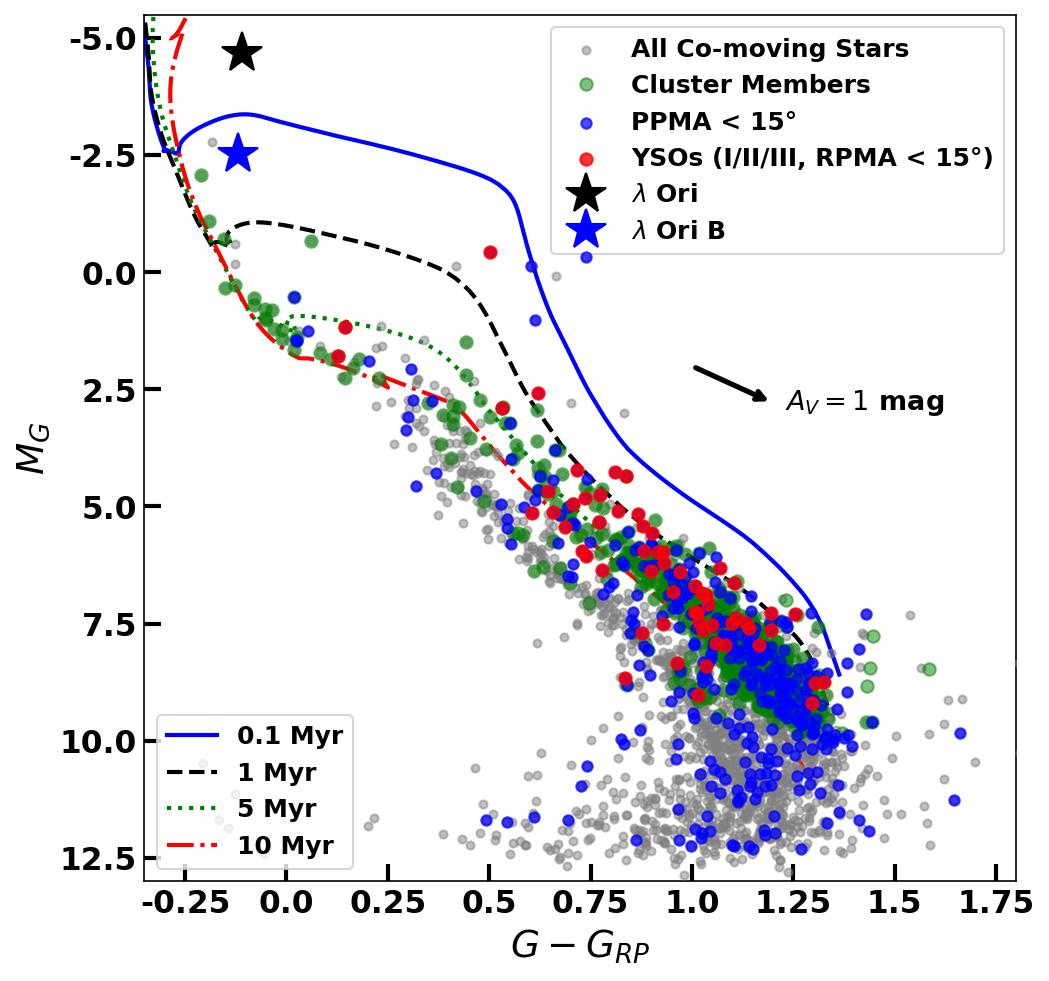}
    \includegraphics[width=\columnwidth]{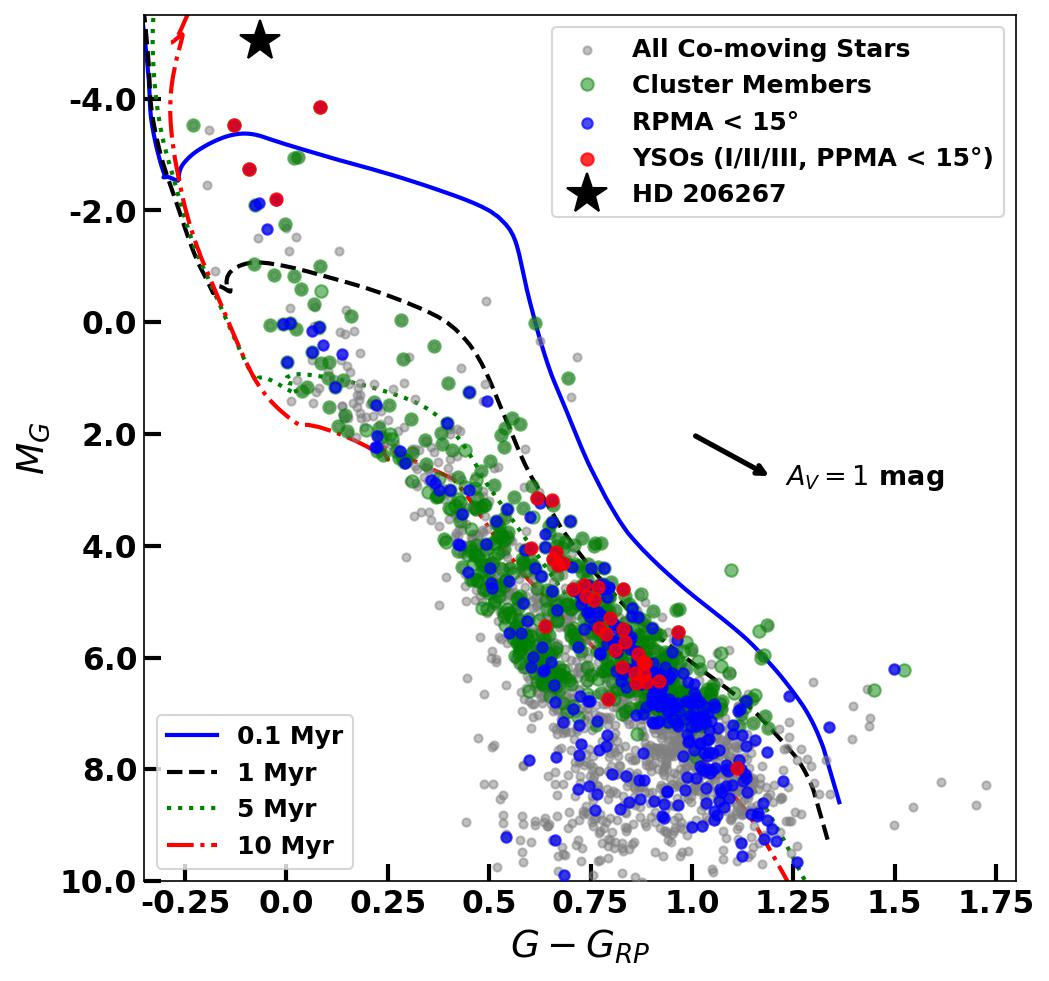}
    \caption{{\it Gaia} EDR3 CMDs of Collinder~69 and IC~1396. 
Grey points show all co-moving stars within the adopted radii, green points mark known cluster members, 
blue points represent stars with RPMA $<$ 15\textdegree, and red points indicate YSOs (Class~II/III) with RPMA $<$ 15\textdegree. 
The positions of the central ionising O-type stars ($\lambda$~Ori, $\lambda$~Ori~B for Collinder~69; HD~206267 for IC~1396) are highlighted with star symbols. 
Isochrones from the MIST models at 0.1, 1, 5, and 10~Myr are overlaid for reference. The black arrow shows the extinction vector corresponding to $A_V=1$ mag.}
    \label{fig:cmd}
\end{figure*}

Small scale sequential star formation and age gradients have been observed in HII regions and BRCs \citep{Sugitani1995ApJ...455L..39S,Maheswar2007MNRAS.379.1237M,Ogura2007PASJ...59..199O,Choudhury2010ApJ...717.1067C}. To examine potential age differences between the radially moving populations (RPMA $< 15^\circ$) and the general cluster members from \cite{Cantat2018A&A...618A..93C}, we constructed {\it Gaia} colour–magnitude diagrams (CMDs) and overlaid Modules for Experiments in Stellar Astrophysics (MESA) isochrones and evolutionary tracks (\href{http://waps.cfa.harvard.edu/MIST/}{MIST})\footnote{http://waps.cfa.harvard.edu/MIST} isochrones at 0.1, 1, 5 and 10 Myr \citep{Choi2016ApJ...823..102C}. \autoref{fig:cmd} shows the {\it Gaia} M\textsubscript{G} vs G–G\textsubscript{RP} CMDs for Collinder 69, and IC 1396. The CMD shows all sources in the astrometric ellipse, all radially moving away sources (RPMA $< 15^\circ$) and Class I/II/III sources with RPMA $< 15^\circ$\ separately. Note that the \citet{Cantat2018A&A...618A..93C} sample has a magnitude limit, which is visible in the y-axis histogram.

We corrected the young clusters for interstellar extinction using the Bayestar19 3D dust map \citep{Green2019ApJ...887...93G} as implemented in the dustmaps package. Using the coordinates and distances of each stars from {\it Gaia}, we extract the reddening value from the dust map. The reddening is converted to {\it Gaia} passbands using the extinction law from \cite{Mathis1990ARA&A..28...37M}, with a total-to-selective extinction ratio, R\textsubscript{v} = 3.1.

Based on the {\it Gaia} CMDs alone, in both clusters, we find no strong evidence for systematic age gradients. Both the radially moving outward population and the broader cluster members largely coincide along similar isochronal ages 4-5 Myrs, indicating that stars currently exhibiting outward motions are not significantly younger than the typical cluster member. However, we caution that optical CMD-based ages of young low-mass pre-main-sequence stars are affected by variability, unresolved multiplicity, accretion, photometric uncertainties, and differential extinction. We also note that this comparison is between kinematically defined subsets of the co-moving population, and not between triggered and non-triggered stars.

In the CMD of Collinder 69 there is a main sequence population of stars inside the astrometric ellipse, which are below the 10 Myr isochrone. The cluster members, the sources with RPMA $< 15^\circ$\ including YSOs are occupying a similar region around 5 Myr isochrone which is the age of the cluster reported in literature \citep{2004barrado, 2011bayo, 2023healy} and show no gradient of age in different populations. IC 1396 also shows the main sequence population, but they are mixed with cluster members from \cite{Cantat2018A&A...618A..93C}. The cluster members are positioned in main sequence population, and near 5 Myr isochrone. We see a higher number of sources with RPMA $< 15^\circ$\ on the 5 Myr isochrone region than the main sequence population. Interestingly, the YSOs are almost completely near 5 Myr isochrone, which is near to the cluster age of 4 Myr \citep{Pelayo2023A&A...669A..22P}. 

It is possible that the cluster selected by \cite{Cantat2018A&A...618A..93C} for IC 1396 has multiple age populations, as reported in previous studies \citep{Sicilia2005AJ....130..188S,Sicilia2014A&A...562A.131S}. We also note that evidence for younger or multiple populations in IC~1396/Tr~37 has not been established from isochrone fitting alone. Previous studies have identified less evolved sources, including flat-spectrum Class~II sources, Class~I objects, and Class~0 protostars associated with dense clouds, globules, and ionisation fronts \citep{Reach2004ApJS..154..385R,Sicilia-Aguilar2006AJ....132.2135S,Sicilia2019A&A...622A.118S}. These embedded and infrared-selected populations provide stronger evidence for recent or ongoing star formation than optical CMD positions alone. But it is interesting to note that the cluster population at 4-5 Myr in the isochrones, and the population with RPMA $< 15^\circ$\ does not show much different ages. It is also noteworthy that the small fraction of cluster members also show younger population of around 1 Myr, indicative of multiple age populations in the overall region.

\noindent The absence of a clear age gradient for the population with RPMA $< 15^\circ$\ with cluster members in both clusters can also be partly attributed to differential extinction, photometric uncertainties, and unresolved multiplicity, and it should not be interpreted as evidence against feedback-driven dynamics. In hierarchical star-forming regions, the superposition of multiple substructures and early-time dispersal can wash out simple spatial age trends \citep{Ward2020MNRAS.495..663W}. Moreover, {\it Gaia} CMDs are less sensitive to embedded protostellar populations, which are better traced using infrared excess, spectral slopes, emission-line diagnostics, or X-ray emission. The REI estimated therefore traces the present-day kinematics of feedback and gas removal, while the CMDs reflect an age-mixed population assembled across a structured medium. Applying more sophisticated extinction corrections and incorporating additional youth indicators, such as IR excess or emission-line features, x-ray emissions, could help clarify whether any age bias exists among the radially moving populations. Because of the above caveats, we use the CMD only as a qualitative comparison to check whether the RPMA-selected population shows a large systematic offset from the broader co-moving population.

\section{Discussion}

The study fits into a broad context of work exploring the interplay of stellar feedback, cluster dynamics, and star formation processes in massive star-forming regions. By leveraging the high-precision astrometry of {\it Gaia} EDR3 \citep{Gaia2016A&A...595A...1G, Lindegren2021A&A...649A...2L}, our detection of a radially outward-moving population of stars strongly supports the notion that the RE  is actively shaping the young stellar environment. This effect, seen previously in individual BRCs \citep[e.g.,][]{Arun2021MNRAS.507..267A,Sahaa2022MNRAS.510.2644S,Saha2022MNRAS.515L..67S}, now emerges as a general kinematic signature in multiple clusters. Our results thus extend the observational evidence for feedback-driven gas dispersal, complementing both theoretical models and simulations that emphasise how ionising photons from massive stars can compress and accelerate surrounding gas, influencing star formation and the subsequent stellar kinematics \citep{Dale2015}.

The absence of a clear age gradient within the radially moving populations challenges the simplest models of sequential star formation triggered by expanding HII regions \citep{Elmegreen1977ApJ...214..725E,Sugitani1991ApJS...77...59S}. While classical scenarios envision newly formed stars appearing systematically farther out along expanding shells of swept-up material, we find that stars currently exhibiting outward motion are not necessarily younger than the general cluster membership. This discrepancy may reflect the complexities noted in recent hydrodynamical simulations, which show that feedback can produce non-monolithic and anisotropic structures of gas and stars \citep{Dale2013MNRAS.430..234D,Girichidis2020SSRv..216...68G}. Rather than generating a simple spatial age sequence, feedback may drive a more chaotic rearrangement, accelerating stars that have already formed in denser pockets without creating a neat temporal progression. Our results resonate with the work of \citet{Kuhn2019ApJ...870...32K}, who identified complex and anisotropic expansion in young clusters, and with \citet{Armstrong2024A&A...692A.166A}, who highlighted direction-dependent dispersal patterns influenced by feedback.

That these radially moving stars often align towards known BRCs and IR bubble rims supports the view that feedback is direction dependent and not uniformly isotropic. Observations by \citet{Sugitani1991ApJS...77...59S,Sugitani1994ApJS...92..163S} and subsequent studies have shown that BRCs mark the edges of ionisation fronts and can serve as potential sites of triggered star formation. However, our RPMA-selected sample is defined kinematically and physical associations with individual BRCs are not investigated. Sources genuinely located at the rims cannot be separated from those seen in projection against them, because the line-of-sight depth of our astrometric selection is $\sim$90 pc for Collinder 69 and $\sim$310 pc for IC 1396, much larger than the thickness of a BRC. This is a long standing difficulty in establishing triggered star formation \citep{Sugitani1991ApJS...77...59S,Choudhury2010ApJ...717.1067C,Getman2012MNRAS.426.2917G,Mookerjea2012A&A...542L..17M,Sicilia2019A&A...622A.118S}, and requires independent tracers of association such as stellar and molecular radial velocities or X-ray and IR youth indicators. We therefore do not interpret the absence of an age offset in the CMDs as evidence about the ages of any triggered population. The lack of a strong age contrast may also stem from observational uncertainties such as unresolved binaries or field star contamination \citep{Bouvier2001A&A...375..989B, DaRio2010ApJ...723..166D, Kounkel2018AJ....156...84K}. Where such information is available, it supports the present picture. \citet{Sicilia2019A&A...622A.118S} combined IRAM molecular line velocities with {\it Gaia} astrometry and found the three dimensional velocity of IC 1396A (SFO 36) to be directed away from Trumpler 37, consistent with the outward motion we infer for this direction.

Our finding that the radially moving subset (RPMA $<15^\circ$) does not exhibit a clear age offset relative to the general cluster membership is consistent with {\it Gaia}-based studies showing that many young, OB-dominated systems are not the
remnants of compact clusters expanding after gas expulsion, but instead reflect hierarchical star formation with significant substructure and subsequent rapid dispersal and possibly galactic shear \citep{Ward2020MNRAS.495..663W}.
In such a framework, anisotropic or preferentially radial proper motions need not be accompanied by a monotonic age sequence: stars form quasi-simultaneously across a clumpy medium and any small age differences are rapidly mixed by the
ambient velocity field. This naturally explains the strong kinematic signal we detect (RPMA excess and REI $>0$) together with the lack of a pronounced age gradient in the {\it Gaia} CMDs.

To obtain a more comprehensive understanding of the region’s star formation history and feedback processes, more detailed and targeted analyses are essential. Spectroscopic follow-up to determine radial velocities and confirm cluster membership will enable the reconstruction of a refined three-dimensional velocity field, facilitating the distinction between genuine members and field interlopers. Near- and mid-IR observations can identify younger protostars and reveal IR-excess sources, potentially establishing whether younger star formation episodes indeed coincide with these feedback-driven motions. Such complementary data may confirm or refute the presence of subtle age gradients currently hidden by observational limitations.

Furthermore, future theoretical work could explore parameter regimes where feedback imprints the kind of kinematic signatures observed here without producing strong age gradients, shedding light on the interplay between ionising radiation, stellar winds, and the local density structure. Recent simulations by \citet{Grudic2021MNRAS.506.2199G} and \citet{Howard2017MNRAS.470.3346H} have shown that the spatial and temporal distribution of star formation in feedback-regulated environments can be quite complex, emphasising that observational tests like ours can give good empirical sign posts.

\section{Conclusions}

In this study, we have examined the kinematic signatures of the RE in two young open clusters, Collinder 69 and IC 1396, which host O-type stars and are embedded inside IR bubbles, using {\it Gaia} EDR3 data. The conclusions drawn in the studies are as follows.

\begin{itemize}

\item Using {\it Gaia} EDR3 astrometric data, we identified an extended population of stars and YSOs with astrometric similarity with the two young clusters.

\item We identified a significant over-density of stars moving radially outward (RPMA $<15^\circ$), quantified using REI,  with its statistical significance confirmed by binomial, Anderson--Darling, and V-tests.
\item Directional analysis reveals that the sources with RPMA $< 15^\circ$~have anisotropic distribution aligning with directions of BRCs, indicative of RE.

\item A comparison of CMDs does not show a clear age gradient between the radially moving stars and the overall cluster membership. 

\item The analysis of both clusters suggests that the RE can contribute to the present-day kinematic structure of star-forming regions. 

\end{itemize}

In summary, our study highlights how O-type stellar feedback and IR bubbles shape cluster kinematics in complex, non-linear ways. We confirm that feedback leaves detectable kinematic traces without necessarily following orderly age sequences. The ROCKETS project will continue this work through deeper analysis and spectroscopic studies of other star-forming regions.
\section*{Acknowledgements}

We thank the anonymous referee for the valuable comments on the manuscript. R.A.\ acknowledge support from the ANID -- Millennium Science Initiative Program -- Center Code NCN2024\_001 and ANID FONDECYT postdoctorado grant 3260521. P.S. was partially supported by a Grant-in-Aid for Scientific Research (KAKENHI No JP24K17100) of the Japan Society for the Promotion of Science (JSPS). This work has made use of data from the European Space Agency (ESA) mission
{\it Gaia} (\url{https://www.cosmos.esa.int/gaia}), processed by the {\it Gaia} Data Processing and Analysis Consortium (DPAC, \url{https://www.cosmos.esa.int/web/gaia/dpac/consortium}). Funding for the DPAC has been provided by national institutions, in particular the institutions participating in the {\it Gaia} Multilateral Agreement. We acknowledge the use of ChatGPT, developed by OpenAI, for assistance with language editing and grammatical corrections. The authors retain full responsibility for the content of this manuscript.

\section*{Data Availability}

The complete results produced by this study will be made available to interested parties upon request to the corresponding author

\bibliographystyle{mnras}
\bibliography{example}

\newpage
\appendix

\section{KDE Distribution}\label{app:kde}

This appendix presents the distribution of REI values for the external open-cluster reference sample described in Sect.~3.1. The sample consists of the 314 clusters common to \cite{Cantat2018A&A...618A..93C} and \cite{Hunt2023A&A...673A.114H} with at least 100 members and distances below 1.5 kpc, for which the RPMA and REI were computed in the same way as for Collinder 69 and IC 1396.

\begin{figure}
    \includegraphics[width=\columnwidth]{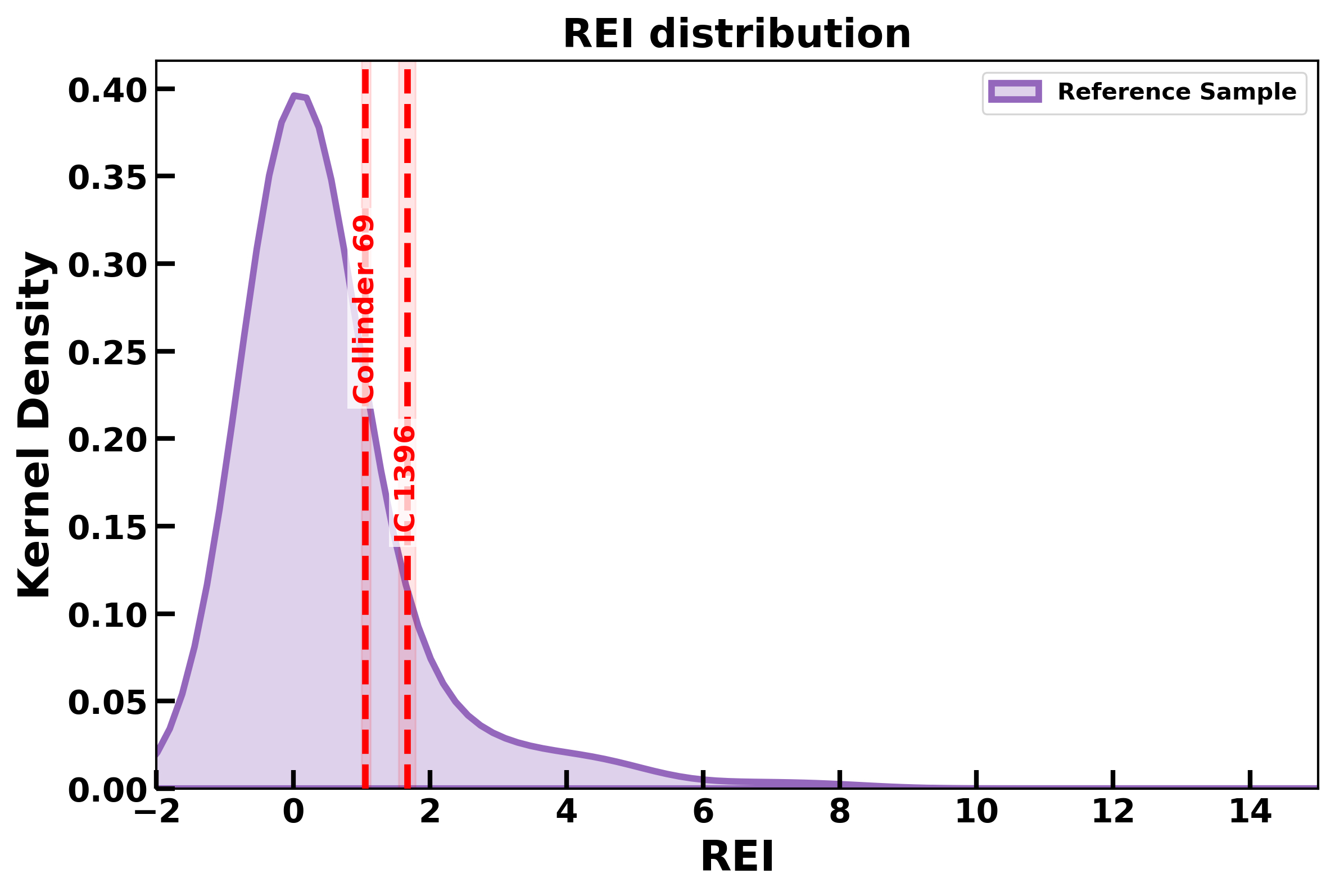}

    \caption{KDE distributions of REI for the external open-cluster reference sample of 314 clusters (purple). The vertical dashed lines correspond to the values for the two young clusters analysed in this study. The REI values for the young clusters are higher than the median value of the reference sample, indicating an enhanced over-density of stars moving radially outward.}  \label{fig:kde}
\end{figure}
\end{document}